\documentclass[11pt,a4paper]{article}
\usepackage{jheppub}
\usepackage{amsmath,amssymb,amsfonts,amsthm}
\usepackage{graphicx}
\usepackage{epsfig} 
\usepackage{verbatim} 
\usepackage{enumerate}
\usepackage{slashed} 
\usepackage{epsf}

\usepackage{todonotes}

\def\stamp{--- {\bf \today} --- {\bf \jobname.tex}}

\def\stamp{--- {\bf \today} --- {\bf \jobname.tex}}

\def\sign(#1){\textrm{sign}(#1)}

\def\BE{\begin{equation}} 
\def\EE{\end{equation}}

 \def\<#1|#2){\left\langle#1|#2\right\rangle} 
 \def\<#1|#2|#3]{\left\langle#1|#2|#3\right ]} 
\def\(#1|#2|#3>{\left[#1|#2|#3\right\rangle} 
 \def\[#1|#2]{\left[#1|#2\right]}

\def\an[#1,#2]{\left\langle#1\,#2\right\rangle} 
\def\aq[#1,#2,#3]{\left\langle#1|#2|#3\right]} 
\def\qa[#1,#2,#3]{\left[#1|#2|#3\right\rangle} 
\def\sq[#1,#2]{\left[#1\,#2\right]} 
\def\spa#1.#2{\left\langle#1\,#2\right\rangle} 
\def\spab[#1,#2,#3]{\left\langle#1|#2|#3\right]} 
\def\spba[#1,#2,#3]{\left[#1|#2|#3\right\rangle} 
\def\spb#1.#2{\left[#1\,#2\right]} 
\def\lor#1.#2{\left(#1\,#2\right)}

\newcommand{\beq}{\begin{eqnarray}}
\newcommand{\eeq}{\end{eqnarray}}

\newcommand{\ba}{\begin{eqnarray}}
\newcommand{\ea}{\end{eqnarray}}

\title{Effective Field Theory and Electroweak Baryogenesis in the Singlet-Extended Standard Model} 
 
 \author[a]{P.~H.~Damgaard,}
\author[b]{A. Haarr,}
\author[c]{D. O'Connell,}
\author[b]{A. Tranberg,}

\affiliation[a]{Niels Bohr International Academy and Discovery Center,\\ 
The Niels Bohr Institute, University of Copenhagen, \\Blegdamsvej 17, 
DK-2100 Copenhagen \O, Denmark,}
\affiliation[b]{Faculty of Science and Technology, University of Stavanger, \\4036 Stavanger, Norway}
\affiliation[c]{Higgs Centre for Theoretical Physics, School of Physics and Astronomy, \\
The University of Edinburgh,
Edinburgh EH9 3JZ, Scotland, UK}

\emailAdd{phdamg@nbi.dk}
\emailAdd{anders.haarr@uis.no}
\emailAdd{anders.tranberg@uis.no}

\preprint{Edinburgh 2015/30}
 
\abstract{Electroweak baryogenesis is a simple and attractive candidate mechanism for generating the observed baryon asymmetry in the Universe. Its viability is sometimes investigated in terms of an effective field theory of the Standard Model involving higher dimension operators. We investigate the validity of such an effective field theory approach to the problem of identifying electroweak phase transitions strong enough for electroweak baryogenesis to be successful. 
We identify and discuss some pitfalls of this approach due to the modest hierarchy between mass scales of heavy degrees or freedom and the Higgs, and the possibility of dimensionful couplings violating the decoupling between light and heavy degrees of freedom.
} 
 
\keywords{Baryogenesis, Standard Model, Singlet Scalar Extension, Electroweak Phase Transition, Effective Field Theory} 
 
\begin{document} 
\maketitle 
 
\section{Introduction}\label{sec:introduction} 

Electroweak baryon number violation in association
with a strongly first-order phase transition 
provides a compelling scenario for explaining
the observed baryon asymmetry of the Universe
\cite{Kuzmin:1985mm} (see, for example, ref. \cite{Rubakov:1996vz} for
a comprehensive review). However, the robust bound
on the Higgs mass needed to ensure a first-order 
finite-temperature electroweak phase transition
\cite{Kajantie:1996mn} is incompatible with the experimental value of $m_H\simeq125$ GeV\footnote{We will take the Higgs mass to be $m_H = 125.7$ GeV below for concreteness.}. In addition, despite initial optimism~\cite{Shaposhnikov:1986jp,Shaposhnikov:1987pf,Farrar:1993sp}, it is now generally believed that the Standard Model
does not provide enough CP violation to obtain the observed baryon number~\cite{Gavela:1993ts,Gavela:1994ds,Gavela:1994dt,TranbergCP}.

Nevertheless, the idea of electroweak baryogenesis is natural and attractive, 
and it is known that new physics
which couples to the Higgs sector can easily induce a first-order transition
\cite{Anderson:1991zb,Profumo:2007wc,Barger:2011vm,Curtin:2012aa,Chung:2012vg,Morrissey:2012db,Cline:2012hg, Damgaard:2013kva, Kozaczuk:2015owa,Huang:2015tdv}. 
Specific models for this new physics have been proposed, but there is a very 
large model space and individual models are constrained by Large Hadron Collider (LHC) data and other observables (see for instance \cite{Donal,Damgaard:2013kva,Katz,Curtin:2014jma, Kozaczuk:2015owa, Fuyuto:2015ida,Arkani-Hamed:2015vfh}). 

It would clearly be advantageous if one could go one level of abstraction higher, and view all possible extensions
of the Standard Model capable of producing the required first-order phase transition in a unified way.
Effective Field Theory (EFT) is a universal language which holds the promise of deriving just such
a model-independent bound
on electroweak baryogenesis. In this approach, physics beyond the Standard Model
is parametrized through a systematic expansion in higher dimension operators that are suppressed
by the new energy scale. Indeed, baryogenesis in the context of effective theory has already been 
explored~\cite{Zhang:1992fs,Grojean:2004xa,Bodeker:2004ws,Grinstein:2008qi,Huang:2015izx}.

Effective field theories expand low energy physics, characterized by an energy scale $E$, in powers of the ratio $E/\Lambda$ where $\Lambda$ is the cutoff of the effective theory. Therefore, the conclusions of any EFT analysis can only be expected to be valid when $E/\Lambda \ll 1$ and when the relevant physical processes are consistent with this perturbative expansion.
In the context of an EFT approach to baryogenesis, we must therefore first investigate whether these conditions hold, taking into account that thermal effects and a change of vacuum structure are phenomena that we must expect to play important roles.

In this paper, we investigate the validity of an effective approach to electroweak baryogenesis in the Standard Model augmented by dimension-six operators with cutoffs of order a few TeV. We compare the phase transition structure in the effective model with a specific model of new physics, namely a singlet scalar field that couples only to the Higgs field of the Standard Model. We observe that the separation of scales is not necessarily large enough to prevent phase transitions from occurring in which the new physics scalar makes a transition. Moreover, the domain of validity of the effective theory covers a comparatively small fraction of the parameter space of the full singlet model (see also a recent analysis by \cite{Gorbahn}). Consequently, the predictions of an EFT analysis of baryogenesis need to be treated with some caution.

We have organized our paper as follows. We open in section \ref{sec:matching} by discussing our candidate model for physics above a few TeV: the Standard Model coupled to a single scalar field. We integrate the singlet out and match parameters to our effective theory, which is simply the Standard Model extended by two dimension-six operators. In section \ref{sec:effectivePT} we compute the one-loop finite temperature effective potential of this EFT, and identify regions in parameter space corresponding to first and second order phase transition. In section \ref{sec:singletPT} we perform the same finite temperature analysis in the singlet-extended Standard Model.  Finally, in section \ref{sec:comparison} we compare the predictions of the effective theory and the full singlet model in the region of parameter space where we expect the EFT to be reliable. We conclude in section \ref{sec:conclusions}. Several technical details of the calculations are relegated to the appendices. 

\section{The singlet extension of the Standard Model: matching to the effective Standard Model.}
\label{sec:matching}

We begin by discussing the singlet extension of the Standard Model with one additional real scalar
field $S$. The scalar is a singlet under all gauge symmetries and it couples
to the Standard Model only through the Higgs portal, where for generality
all possible renormalizable couplings are included. 
The relevant part of the Higgs-singlet Lagrangian is then
\beq
{\cal L} = (D_{\mu}H)^{\dagger}D^{\mu}H + \frac{1}{2}(\partial_{\mu}S)^2
- V(H,S) ,
\eeq
where $D_{\mu}$ is the usual covariant derivative and $H$ denotes the Standard Model Higgs 
doublet. The zero-temperature potential $V(H,S)$ can be parametrized in terms of
8 real parameters $V_0,m^2,\lambda,\delta_1,\delta_2,\kappa_2,\kappa_3$ and $\kappa_4$:
\begin{eqnarray}
V(H,S) &=& V_0+{m^2 \over 2} (H^{\dagger}H) +{\lambda \over 4} (H^{\dagger}H)^2 +
{\delta_1 \over 2}(H^{\dagger}H)S  +{\delta_2 \over 2}(H^{\dagger}H)S^2\cr
&& +{\delta_1 m^2 \over 2 \lambda}S+
{\kappa_2 \over 2} S^2 +{\kappa_3 \over 3}S^3+{\kappa_4 \over 4}S^4. 
\end{eqnarray}
The coefficient of the term linear in $S$ has been chosen to ensure
that $S$ has zero vacuum expectation value at tree level. Note that this choice is not a constraint: one may always redefine $S$ by a finite shift to achieve this. Since we have chosen $\langle S\rangle=0$ in the vacuum, the coupling $\delta_2$ does not provide any mixing at zero temperature, although it may do so at finite temperature. Meanwhile, the parameter $\delta_1$ does not vanish in general, and the $S$ and Higgs particles may therefore mix at zero temperature. This phenomenon could provide an experimental signal of the existence of such a singlet. 

The theory has two mass eigenstates, one of which must, of course, correspond to the observed state at roughly 125.7 GeV, which we denote by $\xi_h$. In view of the current success of the Standard Model description of this state, we assume that the 125.7 GeV state is mostly Higgs. As in ref.~\cite{Damgaard:2013kva}, we are mostly
interested in the situation where the other mass eigenstate, denoted by $\xi_S$ (``mostly $S$") is heavier than the observed Higgs particle.
If effective field theory is to be a good description for such a situation, this heavier state must eventually decouple, leaving as its only trace higher dimensional Higgs operators suppressed by the large mass scale. 
To determine how this occurs, it is useful to explicitly diagonalise the mass matrix.
We choose unitary gauge and expand the remaining 
single real component of the Higgs doublet about 
\beq
H(x) ~=~ \frac{1}{\sqrt{2}}(\phi + h(x)) ,
\eeq
where, in the vacuum,
\beq
\langle \phi+h\rangle =v ~\equiv~ \sqrt{\frac{-2m^2}{\lambda}}.
\eeq
The mass matrix is read off from
\beq
\label{eq:vmass}
V_{ \rm mass}={1 \over 2}\left(\mu_h^2 h^2 +\mu_S^2 S^2 +\frac{\mu_{hS}^2+\mu_{Sh}^2}{2} h S \right),
\eeq
with coefficients
\beq
\mu_h^2=\frac{m^2}{2}+\frac{3\lambda v^2}{4}=-m^2~,
~~~ \mu_S^2= \kappa_2+\frac{\delta_2 v^2}{2} ~,~~~\mu_{hS}^2=\mu_{Sh}^2=\delta_1 v ~.
\label{massmatrix}
\eeq
The two mass eigenstates $\xi_h$ and $\xi_s$ therefore have masses given by
\begin{eqnarray}
m_{h}^2 &=& \frac{\mu_h^2 + \mu_S^2}{2} + \frac{\mu_h^2 - \mu_S^2}{2} \sqrt{1 + x^2}, \cr
m_S^2 &=& \frac{\mu_h^2 + \mu_S^2}{2} - \frac{\mu_h^2 - \mu_S^2}{2} \sqrt{1 + x^2},
\end{eqnarray}
where $x \equiv \mu_{hS}^2/ (\mu_h^2 -\mu_S^2)$ sets the strength of mixing. Following
the notation of ref. \cite{Damgaard:2013kva}, we define the mixing angle $\theta$
through
\begin{align}
\xi_s &= \cos \theta \; s - \sin \theta \; h, \\
\xi_h &= \sin \theta \; s + \cos \theta \; h.
\end{align}
where the mixing angle is determined by
\begin{equation}
\tan \theta = \frac{x}{1 + \sqrt{1+x^2}}.
\end{equation}
Thus, as $\mu_S \to \infty$, for fixed $\mu_{hS}$ and $\mu_h$, the mixing angle goes to zero, and the two
states decouple. As expected, this mixing is therefore not an obstacle towards 
performing a large mass expansion (in $1/\mu_S$) and deriving the corresponding effective field theory. 
Nevertheless, in detail the mixing does introduce a few subtleties that we will
describe below.

For our purposes, it is convenient to go to the large $\mu_S$ region by taking $\kappa_2$ large. Provided that $\kappa_2$ is the only large mass scale, we see that the physical mass (squared) of the mostly $S$ scalar is $m_S^2 \simeq \kappa_2 \simeq \mu_S^2$.
To compare with the effective theory, we expand both the $\xi_h$ mass and its 
$\cos \theta$ mixing factor. The relevant expressions are
\begin{align}
\label{eq:mass}
m^2_h &= \mu_h^2 - \frac{\mu_{hs}^4}{4 \kappa_2} + \frac{\delta_2 v^2 \mu_{hs}^4 - 
2 \mu_{hs}^4 \mu_h^2}{8 \kappa_2^2} + O\left(\frac{1}{\kappa_2^3} \right), \\
\cos \theta &= 1 - \frac{\mu_{hs}^4}{8 \kappa_2^2} + O\left(\frac{1}{\kappa_2^3} \right).
\label{eq:mix}
\end{align}
In the $\kappa_2\rightarrow\infty$ limit, the mixing angle goes to zero and $m_h^2\rightarrow \mu_h^2$, as expected.

\subsection{Integrating out the singlet at tree level}
\label{sec:integrateout}

Let us now integrate out the $S$ field\footnote{Note that we do not integrate out the mass eigenstate $\xi_S$, but the field $S$.} and find the effective action
for the $H$ field. For simplicity, we concentrate on the case where $\kappa_3$ and $\kappa_4$
vanish. The action is then quadratic in the $S$, and we can integrate it
out exactly. We write this $S$ part of the Lagrangian as
\begin{align}
\mathcal{L}_S &= - \frac{1}{2} S \partial^2 S - \frac{\delta_1}{2} \left( H^\dagger H 
-\frac{v^2}{2} \right) S - \frac 12 \left( \kappa_2 + \delta_2 H^\dagger H\right) S^2. 
\end{align}
Working at tree level, integrating out replaces the $S$ Lagrangian with 
\begin{equation}
\mathcal{L}_S \rightarrow \frac{\delta_1^2}{8 \kappa_2} \left( H^\dagger H -\frac{v^2}{2} \right) \frac{1}{1 + \frac{\partial^2}{\kappa_2} + 
\frac{1}{\kappa_2} \delta_2 H^\dagger H} \left( H^\dagger H -\frac{v^2}{2}\right),
\label{treelevelSEFT}
\end{equation}
where we understand the differential operator in the denominator by its perturbative expansion.
Expanding to leading order in $1/\kappa_2$, and performing one partial integration, the Lagrangian becomes
\begin{equation}
\label{eq:singlet_expanded}
\mathcal{L}_S \rightarrow
\frac{\delta_1^2}{8 \kappa_2^2} \partial_\mu\left( H^\dagger H\right) \partial^\mu\left( H^\dagger H\right)
-\left(\frac{\delta_1^2\delta_2}{16 \kappa_2^2v^2}-\frac{\delta_1^2}{8 \kappa_2}\right)\left( H^\dagger H -\frac{v^2}{2}\right) ^2
+\frac{\delta_1^2\delta_2}{8 \kappa_2^2} \left( H^\dagger H -\frac{v^2}{2}\right) ^3 .
\end{equation}
Thus, integrating the $S$ out (at tree level) leads to the addition of two dimension 6 operators to the effective $H$ Lagrangian, in addition to shifting the quartic coupling of the $H$. It is straightforward to determine the effects of the terms ${\kappa_3 \over 3}S^3+{\kappa_4 \over 4}S^4$ in the $S$ potential; these operators lead to additional terms in the effective action which are suppressed by higher powers of $\kappa_2$. We will take $\kappa_2$ large, and discard terms throughout which are subleading in $1/\kappa_2$.

\subsection{Integrating out the singlet at one-loop order}
\label{sec:oneloop}

The tree-level contributions to the effective Lagrangian, Eq.~\eqref{eq:singlet_expanded}, are expected to be dominant in most situations. However, if one imposes a $Z_2$ symmetry $S \rightarrow -S$ on the model, then $\delta_1 = 0$. In that case, these tree contributions vanish. Since we will be interested in the $Z_2$ symmetric case below, we compute one-loop corrections to the effective action in the $Z_2$ case. To find the one-loop contribution to the operators $(H^\dagger H)^3$ two equivalent approaches are available: one may either compute the full Coleman-Weinberg contribution to the effective potential and expand it to the appropriate order, or one may perform the actual loop integral. The coefficient of $\partial_\mu(H^\dagger H)\partial^\mu(H^\dagger H)$ is also easily found using the second option. The result can be phrased as contributions to the effective action, and reads
\ba
\delta\Gamma_{\rm eff}=\int d^4x\bigg[\frac{\delta_2^2}{24(4\pi)^2\kappa_2}\partial_\mu(H^\dagger H)\partial^\mu(H^\dagger H)\bigg]
-\int d^4 x\bigg[\frac{\delta_2^3}{12(4\pi)^2\kappa_2} (H^\dagger H)^3\bigg].
\label{eq:loopmatch}
\ea

\subsection{The dimension-6 extended Standard Model}
\label{sec:effective}

Given the results of the previous sections, we can now map the singlet-extended Standard Model to an effective field theory, consisting of the Standard Model augmented by
any higher dimension operators that are allowed by symmetry. Truncating at dimension-six operators,
we can take the Higgs field part of the effective field theory action to be
\begin{equation}
\mathcal{L} = \partial_\mu H^\dagger \partial^\mu H + 
\frac{\delta Z}{v^2} \partial_\mu (H^\dagger H) \partial^\mu (H^\dagger H) - 
\frac{\lambda_4}{4} \left(H^\dagger H - \frac{v^2}{2} \right)^2 - 
\frac{\lambda_6}{\Lambda^2} \left(H^\dagger H - \frac{v^2}{2} \right)^3 ,
\label{eq:eft}
\end{equation}
in terms of two new parameters $\delta Z$ and $\lambda_6$. We have
chosen the coefficient of the $\delta Z$-term to simplify the formulae in
what follows. The last term, of scaling dimension six, is suppressed by $1/\Lambda^2$ 
(and will be power-counted accordingly.) The trade-off in scales is 
incorporated into $\delta Z$. We will throughout take $\Lambda$ to be 1 TeV. Of course, there
are many more dimension-six operators, but the ones shown are those relevant for the effective potential in the $H$-field and those we need to analyze in order to search for phase transitions.

Comparing directly to (\ref{eq:singlet_expanded}), we find at tree level
\beq
\label{eq:match_par}
\frac{\delta Z}{v^2} = \frac{\delta_1^2}{8 \kappa_2^2} , \qquad
\frac{\lambda_6}{\Lambda^2} =  \frac{\delta_1^2 \delta_2}{8 \kappa_2^2}.
\eeq
Meanwhile, working at one-loop order in the $Z_2$ symmetric case, we find using Eq.~\eqref{eq:loopmatch}
\beq
\label{eq:match_par_loop}
\frac{\delta Z}{v^2}  = +\frac{\delta_2^2}{24 (4\pi)^2 \kappa_2}, \qquad
\frac{\lambda_6}{\Lambda^2} = +\frac{\delta_2^3}{12 (4 \pi)^2 \kappa_2} .
\eeq
We have arranged the Lagrangian in Eq.~\eqref{eq:eft} such that the physical scalar mass is independent of $\lambda_6$. We will use this fact below to determine $\lambda_4$ in terms of the measured 125.7 GeV scalar mass.
In a similar way, there is an implicit matching of the coefficient of $H^\dagger H$, absorbed in the requirement that $v$ should have the correct value. 

These expressions allow us to relate the parameter space in the singlet model to the parameter space in the effective model. As usual, the relation between the full singlet model and the EFT can be thought of as a projection. For example, to the order we work at, all values of $\kappa_{3,4}$ are mapped to the same values of $\delta Z$ and $\lambda_6$, which are in turn expressions in three parameters $\delta_1,\delta_2$ and $\kappa_2$. 

Expanding the Higgs doublet in unitary gauge
we see that because of the effective derivative term (proportional to $\delta Z$), the combination
\begin{equation}
\tilde h = (1 + \delta Z) h,
\label{normalization}
\end{equation}
is the canonically normalized field, rather than $h$ itself.
Thus, there is a mixing factor $(1 - \delta Z)$ at all $\tilde h$ interaction vertices.
The pole mass of the $\tilde h$-field is
\begin{equation}
\tilde m^2 = \frac{\lambda_4v^2}{2}  (1 - 2 \delta Z)\equiv m_h^2,
\label{higgsmass}
\end{equation}
and this is the physical mass (125.7 GeV) of the state that is the Higgs particle. We will use this relation to eliminate $\lambda_4$ in the effective theory. 

\section{The phase diagram of the effective theory}
\label{sec:effectivePT}

\begin{figure}
\begin{center}
\includegraphics[width=7cm]{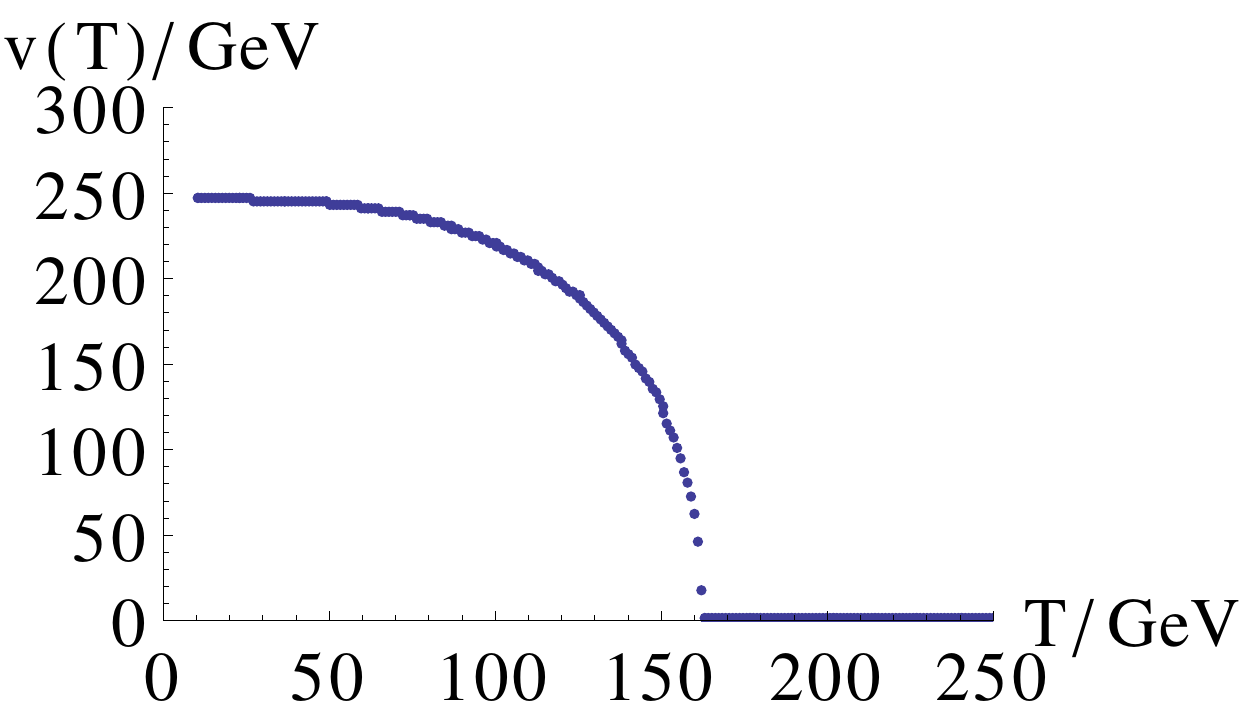}
\includegraphics[width=7cm]{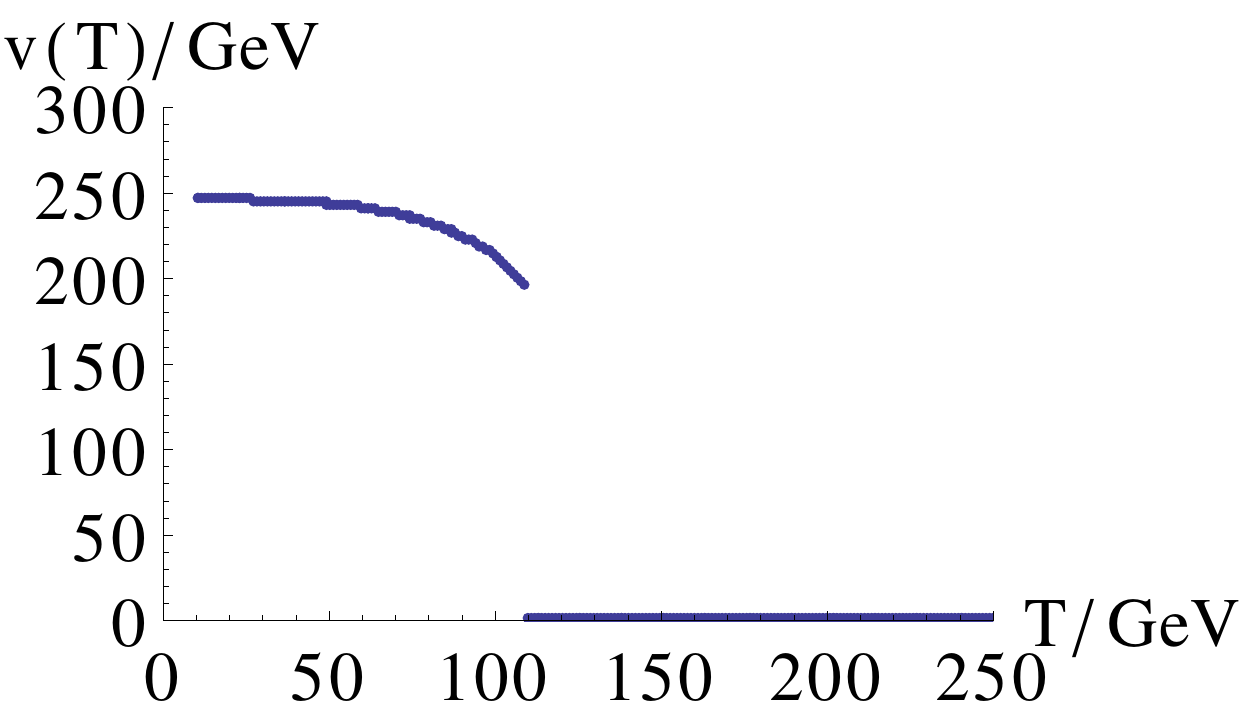}
\caption{Examples of the order parameter dependence on temperature for the effective model. On the left, a second order transition ($T_c\simeq 160$ GeV). On the right a first order transition ($T_c\simeq 105$ GeV).}
\label{fig:ex_vT}
\end{center}
\end{figure}

\begin{figure}
\begin{center}
\includegraphics[width=14cm]{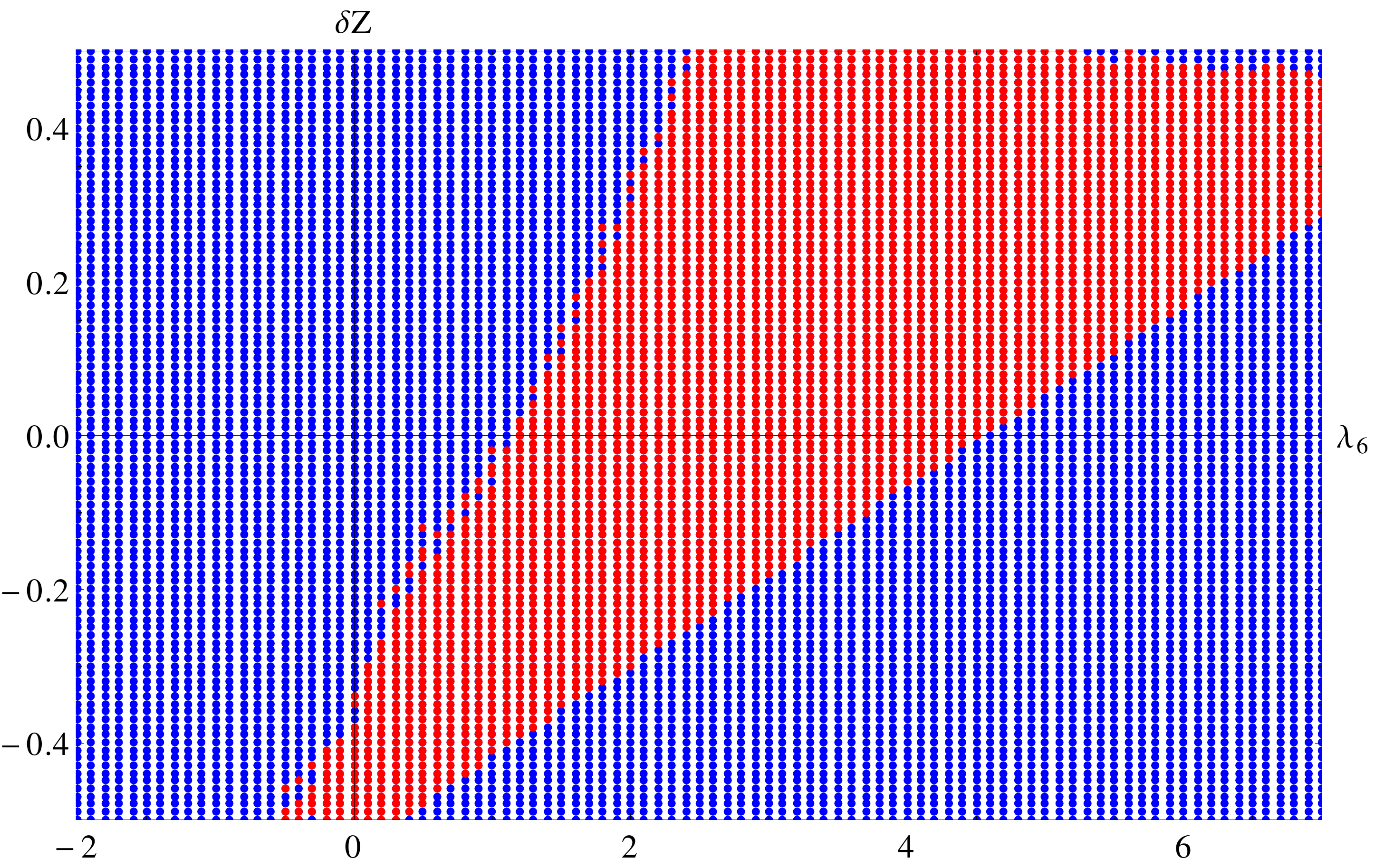}
\caption{Scan of the effective model parameter space ($\delta Z$, $\lambda_6$), identifying 1st order (red) and 2nd order (blue) regions.}
\label{fig:eff_ranges}
\end{center}
\end{figure}

The Higgs potential at finite temperature can be written:
\begin{align}
V_{\rm All}(T) = V+ V_{CW} + V_{ct} + V_T(T),
\end{align}
where $V$ is the tree-level (classical) potential, $V_{CW}$ is the Coleman-Weinberg potential encoding the zero-temperature radiative corrections, $V_{ct}$ are the finite parts of the counterterms that renormalize the theory, and $V_T$ are the corrections due to finite-temperature effects. Appendix \ref{app:finiteT_EFT} contains more explicit expressions for these objects.

We fix the finite counterterms of $V_{ct}$ in such a way that at $T=0$, the minimum of the potential and the mass in the minimum agree with the tree-level values.
\beq
\frac{\partial (V_{ct} + V_{CW})}{\partial H^\dagger} \bigg{|}_{v} = 0, \qquad \frac{\partial^2 (V_{ct} + V_{CW})}{\partial H\partial H^\dagger} \bigg{|}_{v} = 0.
\eeq
Hence, choosing a set of parameters $m_h$, $v$ (or $\lambda$, $m^2$) and $\delta Z$, $\lambda_6$, the complete potential at $T=0$ has the physical vacuum expectation value and physical Higgs mass.
Note that no counterterms for $\delta Z$ and $\lambda_6$ are required at this loop order. 

For a given such set of parameters, the task is then to compute the global minimum of the effective finite temperature potential. This defines the finite temperature ground state, and hence the value of the expectation value $v(T)$, which we take to be the order parameter of the phase transition. In Fig.~\ref{fig:ex_vT} (right) we see an example where $v(T)$ changes from $v=0$ at high temperature to $v=246$ GeV at zero temperature. The curve is continuous but not differentiable at a point, characteristic of a second order phase transition. The critical temperature $T_c$ is defined to be at this cusp. In Fig.~\ref{fig:ex_vT} (left), we see an example of a discontinuous temperature dependence, and hence a first order phase transition. The specific parameters of these examples are
\ba
\delta Z=0.1,&\quad \lambda_6=0.5,&\quad \textrm{ (Second order)},\\
\delta Z=0.1,&\quad \lambda_6=3.0,&\quad \textrm{ (First order)}.
\ea
If the discontinuous ``jump" $\Delta v$ at the critical temperature $T_c$ satisfies
\ba
\frac{\Delta v}{T_c}>0.5,
\ea
we define the transition to be ``strong enough" to provide for baryogenesis \cite{Ahriche:2007jp,sphaleroncriterion,Fuyuto:2014yia}. This is the case for the first order example in the figure, for which this ratio is roughy $1.85$. Clearly, such an extension of the Standard Model with effective operators does provide strong transitions, even when fixing the Higgs mass to the physical value. 

In Fig.~\ref{fig:eff_ranges} we show a scan of a broad parameter range of the effective model. Blue points correspond to second order transitions. Red points are first order transitions. We see that there is a well-defined wedge-shaped region of first order transitions, which is clearly separated from the pure Standard Model point at $(0,0)$ (which is blue). For values of $\delta Z$ less than $-0.5$ and larger than $0.5$, pathological behaviour sets in, corresponding to couplings becoming negative (see for instance Eq.~(\ref{higgsmass})). There seems to be no reason to proceed to sample further negative values of $\lambda_6$. It is possible to proceed to larger values of $\delta Z$, but then the dimension six term is no longer a perturbation to the Standard Model. A similar argument applies to larger values of $\lambda_6$. 

We have tested relaxing the criterion of strength of the phase transition to $0.3$ and even down to $0.1$, leading to only a small shift in the phase boundary. 
Having determined the region in this effective field theory where first order phase transitions can take place, let us move on to address the same questions in the context of the full singlet model.

\section{The phase diagram of the singlet model}
\label{sec:singletPT}

In a completely analogous way, we now proceed to calculate the finite-temperature effective potential in what we take to be the fundamental (UV) theory: the singlet-extended Standard Model. The singlet only couples to the Higgs field and itself as already described in section \ref{sec:matching}. Going to unitary gauge and writing $H=\frac{1}{\sqrt{2}}(\phi+h)$ and trivially setting $S=s$, we have
\beq
V&=&V_0+\frac{1}{4}m^2(\phi+h)^2 + \frac{1}{16}\lambda (\phi+h)^4 + \frac{1}{4}\delta_1(\phi+h)^2s+\frac{1}{4}\delta_2 (\phi+h)^2s^2 \nonumber\\&&+\frac{\delta_1m^2}{2\lambda}s+ \frac{1}{2}\kappa_2 s^2+\frac{1}{3}\kappa_3s^3 + \frac{1}{4}\kappa_4 s^4 .
\eeq
We again introduce the vacuum expectation values
\beq
\langle\phi+h\rangle =v ~\equiv~ \sqrt{\frac{-2m^2}{\lambda}},\qquad \langle s\rangle=0,
\eeq
where the second equation follows from our definition of $s$. At zero temperature,
\beq
\frac{dV}{d\phi}|_{v,0}=\frac{dV}{ds}|_{v,0}=0.
\eeq
The mass matrix in that minimum reads
\beq
M_{\rm min}=\left(\begin{array}{cc}
\frac{d^2V}{dh^2}&\frac{d^2V}{dhds}\\
\frac{d^2 V}{dsdh}&\frac{d^2V}{ds^2}
\end{array}\right)|_{v,0}
=\left(\begin{array}{cc}
\mu_h^2&\frac{\mu_{Sh}^2}{2}\\
\frac{\mu_{hS}^2}{2}&\mu_S^2
\end{array}\right)=
\left(\begin{array}{cc}
-m^2&
\frac{\delta_1}{2}v
\\
\frac{\delta_1}{2}v&
\kappa_2+\frac{\delta_2}{2}v^2
\end{array}
\right),
\eeq
as in (\ref{eq:vmass}). The eigenvalues $m^2_h$ and $m^2_S$ of this mass matrix correspond to the mostly Higgs and mostly singlet mass eigenstates.
This defines an 7-dimensional parameter space
\beq
\{v,\mu_h^2, \mu_{Sh}^2,\mu_S^2,\delta_2,\kappa_3,\kappa_4\};
\eeq
these determine values for $\lambda,m^2,\kappa_2$ and $\delta_1$. Since we know one of the mass eigenvalues and the Higgs vacuum expectation value, let us instead use the equivalent parameters
\beq
\{v,m_S, \theta,m_h,\delta_2,\kappa_3,\kappa_4\},
\eeq
where $\theta$ is the mixing angle. 
From experiment, we set
\beq
m_{h}=125.7\,\textrm{GeV},\qquad v=246\,\textrm{GeV},
\eeq
leaving 
\beq
\{m_S,\theta,\delta_2,\kappa_3,\kappa_4\}
\eeq
to be scanned over. 

The scalar field potential at finite temperature reads
\beq
V(T)= V + V_{\rm CW}+ V_{\rm ct}+ V_{T}(T),
\eeq
where the three remaining components $V_{\rm CW}$, $V_{\rm ct}$ and $V_{T}$ are written out explicitly in Appendix \ref{app:singleteff}. We fix the finite parts of the counterterms so that all first, second, third and fourth derivatives of the effective potential in the zero temperature minimum match the tree-level potential.
For simplicity, we will restrict our sweep to the parameter ranges
\ba
0<\{m_S,\kappa_3\}<4\textrm{ TeV},\quad 
0<\sin\theta <1,\quad
0<\{\delta_2,\kappa_4\}<2\pi.
\ea
In particular, we can specialize to a $Z_2$ symmetric potential ($S\leftrightarrow -S$ is a symmetry), in which case $\sin\theta=\kappa_3=0$ ($\delta_1=0$), leaving only $m_S,\delta_2$ and $\kappa_4$. Note that we do not allow for spontaneous symmetry breaking of $S$, hence $\langle S\rangle=0$ {\it always} when $Z_2$ symmetry is imposed, even at finite temperature.

\subsection{Identifying first and second order transitions in the singlet-extended model}
\label{sec:12singlet}

\begin{figure}
\begin{center}
\includegraphics[width=7cm]{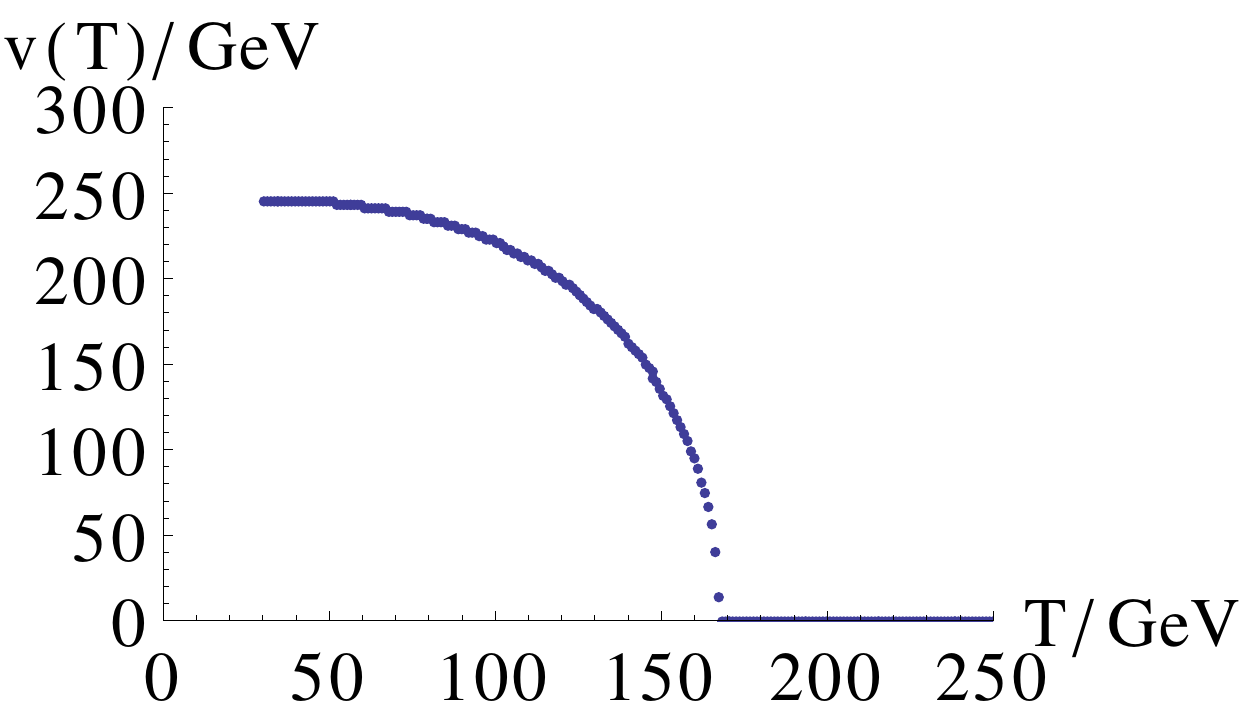}
\includegraphics[width=7cm]{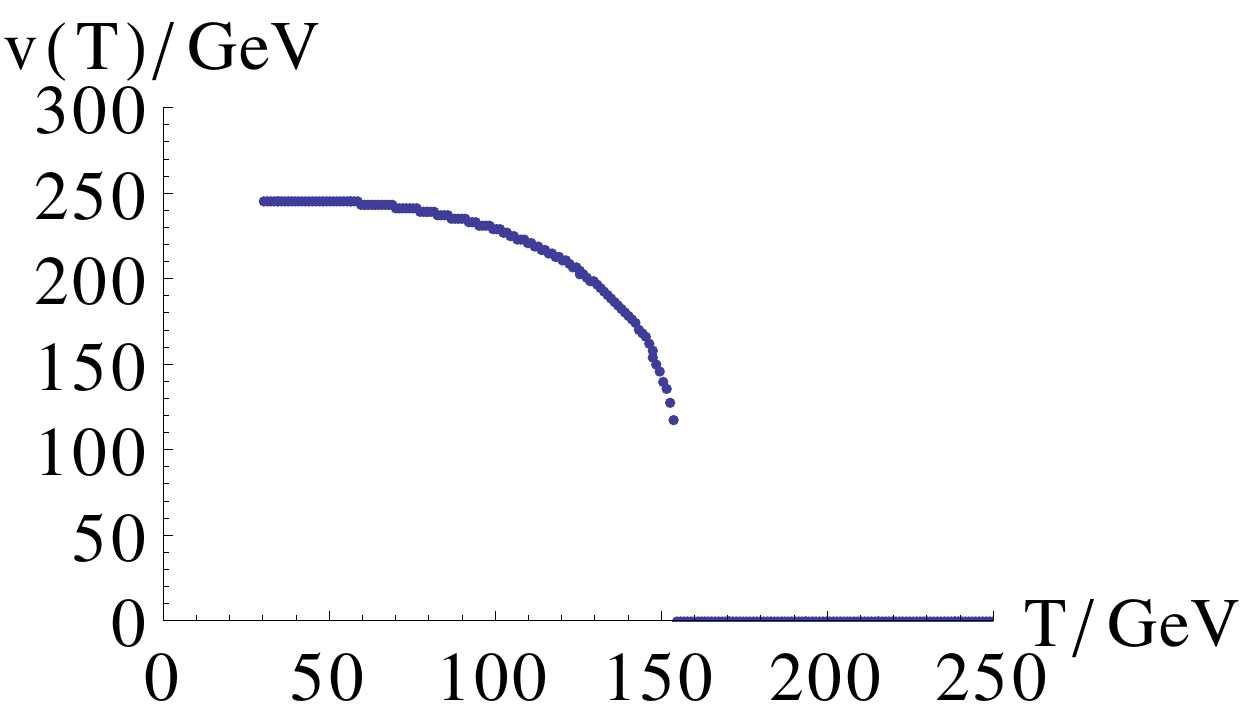}
\caption{Examples of the order parameter dependence on temperature for the singlet extended model with $Z_2$ symmetry. On the left, a second order transition ($T_c\simeq 168$ GeV). On the right a first order transition ($T_c\simeq 155$ GeV).}
\label{fig:ex_vT_sing}
\end{center}
\end{figure}
In the $Z_2$ symmetric case, the order parameter is the Higgs expectation value, and the phase transition may be identified exactly as in the EFT case. Fig.~\ref{fig:ex_vT_sing} shows again examples of first and second order transitions in terms of the temperature dependence of the Higgs expectation value $v(T)$. The specific parameters of these example points are
\ba
&&m_S=2544\textrm{ GeV},\quad \delta_2=6.06,\quad \kappa_4=3.04,\quad \textrm{ (Second order)},\\
&&m_S=378\textrm{ GeV},\quad \delta_2=5.36,\quad \kappa_4=4.95,\quad \textrm{ (First order)/}
\ea
The first order transition has $\frac{v(T_c)}{T_c}=0.75$, so we consider the transition to be strong enough that this point is a candidate for baryogenesis.

\begin{figure}
\begin{center}
\includegraphics[width=7cm]{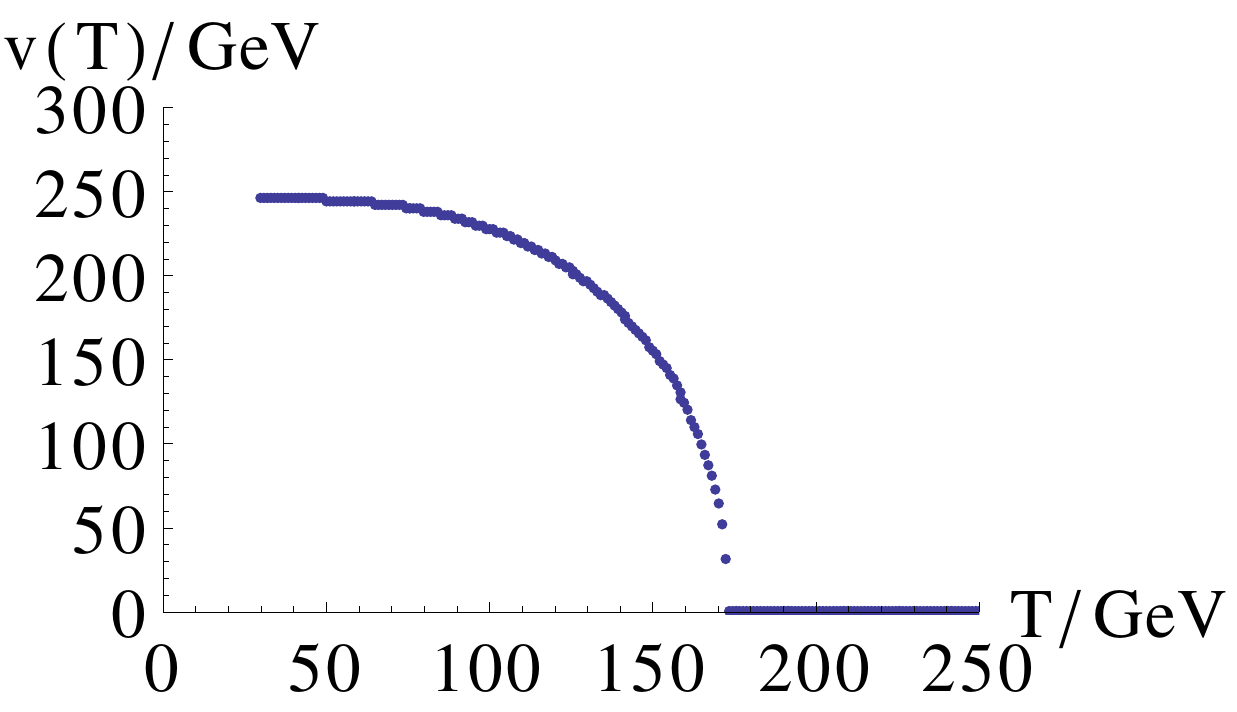}
\includegraphics[width=7cm]{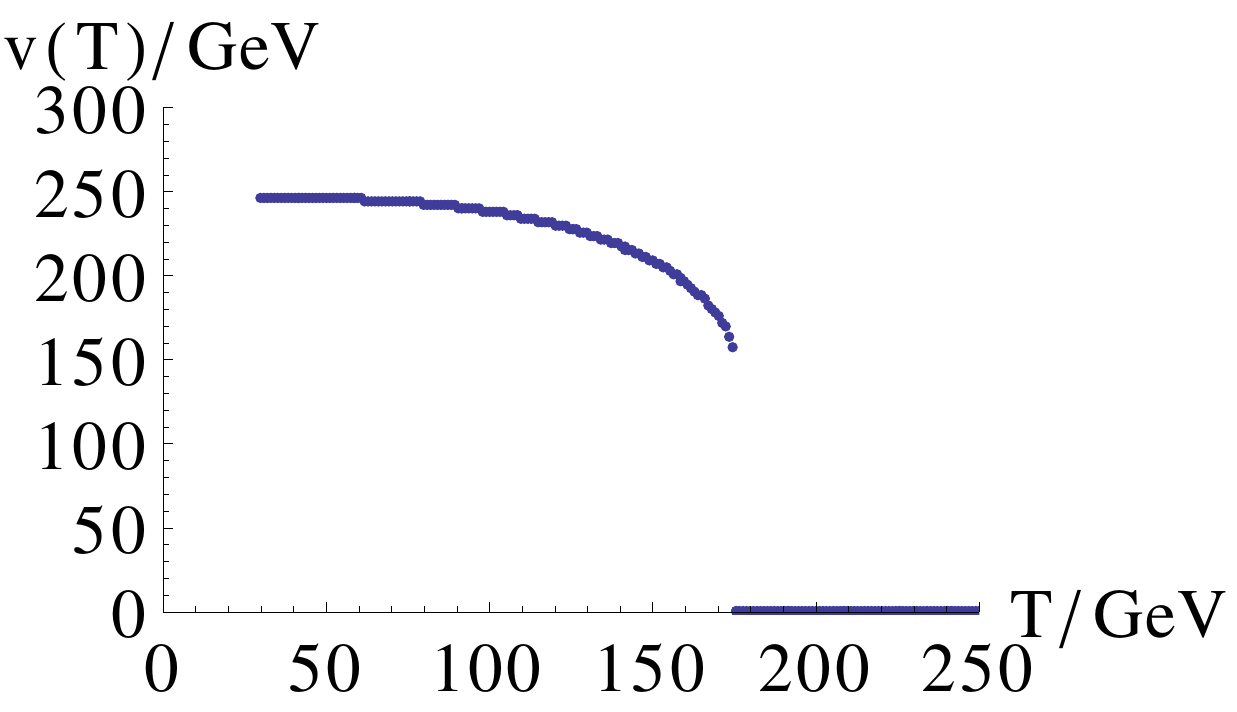}
\includegraphics[width=7cm]{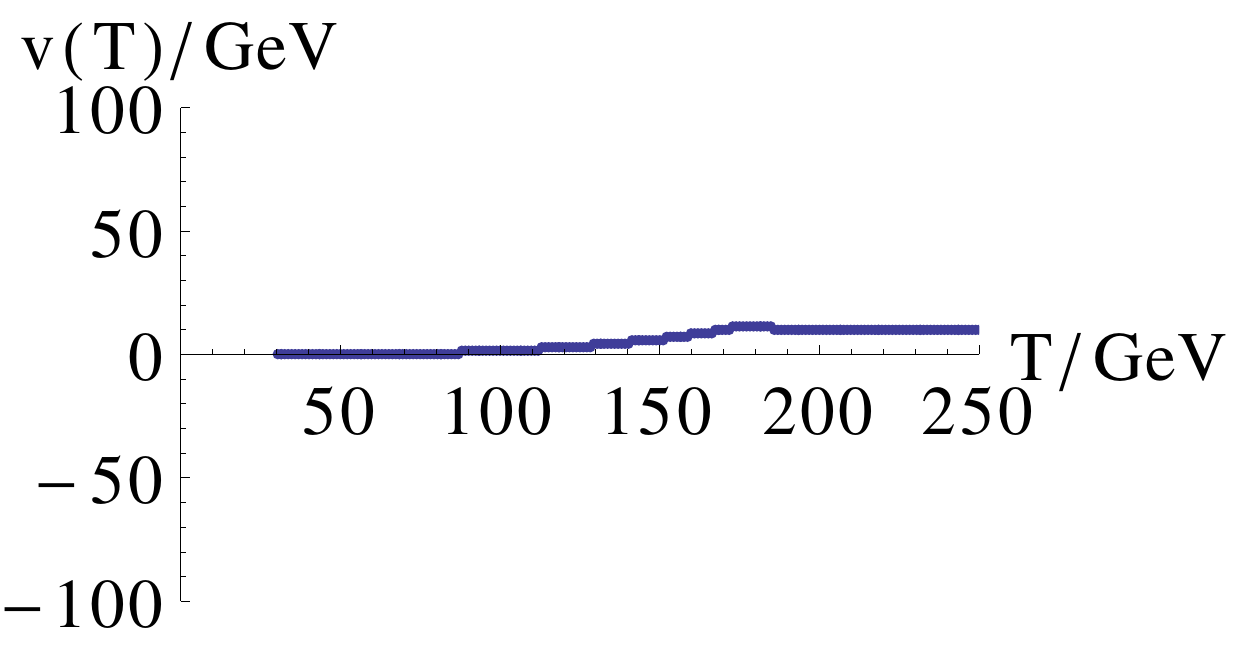}
\includegraphics[width=7cm]{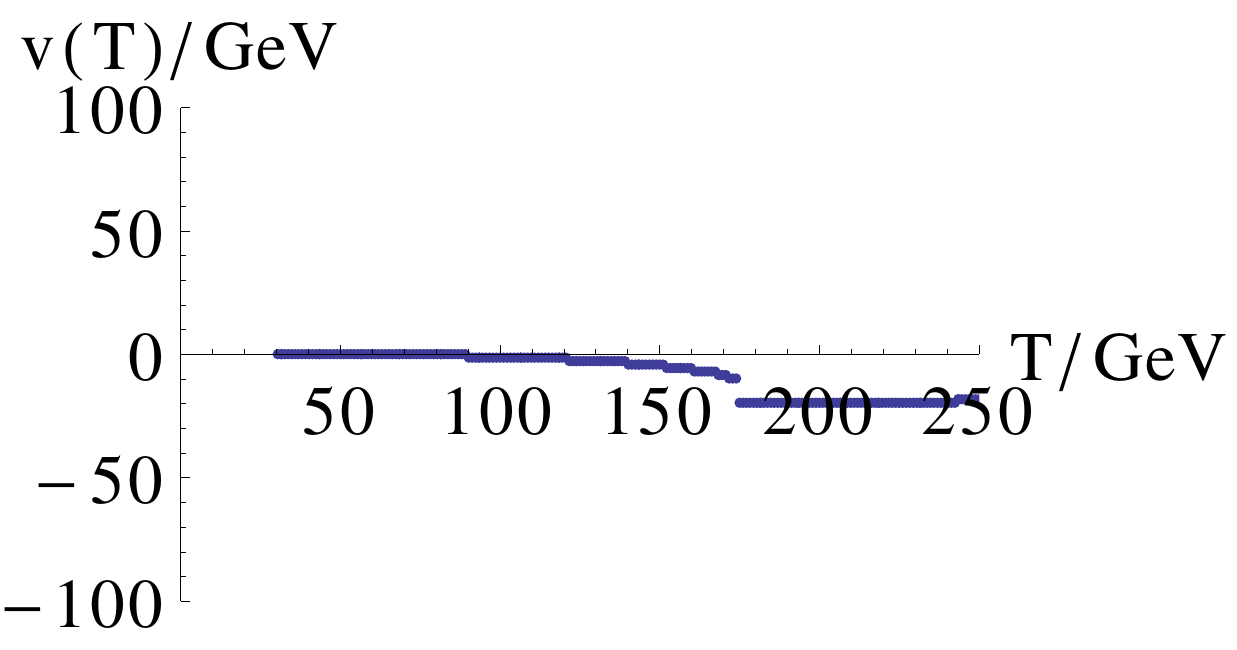}
\caption{Examples of the Higgs (top) and singlet (bottom) finite temperature expectation value for the singlet extended model. On the left, a second order transition ($T_c\simeq 163$ GeV). On the right, a first order transition ($T_c\simeq 165$ GeV).}
\label{fig:ex_sT_sing}
\end{center}
\end{figure}

When not imposing $Z_2$ symmetry, the singlet field also picks up an expectation value $s(T)$ at finite temperature, as shown in Fig.~\ref{fig:ex_sT_sing}. A single order parameter could be chosen to be the combination $\sqrt{s^2(T)+v^2(T)}$, but we will simply consider the two separately and retain only those phase transitions for which the discontinuity in the Higgs-only potential satisfies $\frac{\Delta v}{T_c}>0.5$.

It turns out that quite a number of parameter sets, although they do have a local minimum at $\langle s\rangle=0$, $\langle\phi+h\rangle=v$ as specified by the renormalisation conditions, in fact have a global minimum at $\langle\phi+h\rangle=0$, $\langle s\rangle=\omega\neq 0$.  As discussed previously, one may always shift the field $s$ so that the singlet vacuum expectation value vanishes, but not that of the Higgs field (shifting the singlet $s\rightarrow \omega+\bar{s}$ changes the parameters in such way that the coefficient of $H^\dagger H$ is positive around the global minimum). Since the global minimum has no Higgs vacuum expectation value, we take the corresponding parameter set to be unphysical (or rather, ruled out by the experimental observation of a non-zero Higgs vacuum expectation value). 

In Fig.~\ref{fig:ex_sT_sing} we see examples of first and second order transitions for general parameter sets
\ba
&&m_S=1084\textrm{ GeV},\quad \delta_2=1.68,\quad \theta =0.092,\quad \kappa_3=454\textrm{ GeV},\quad \kappa_4=1.95,\quad \textrm{ (Second order)},\nonumber\\
&&m_S=2063\textrm{ GeV},\quad \delta_2=4.63,\quad \theta =0.146,\quad \kappa_3=1847\textrm{ GeV},\quad \kappa_4=3.08,\quad \textrm{ (First order).}\nonumber\\
\ea
Again, since the first order transition has $\frac{v(T_c)}{T_c}=0.96$, we take it to be strong enough for baryogenesis. Therefore, adding a singlet to the Standard Model may provide a strong first order transition, with or without $Z_2$-symmetry.

\begin{figure}[t]
	\centering
	\includegraphics[width=0.7\textwidth]{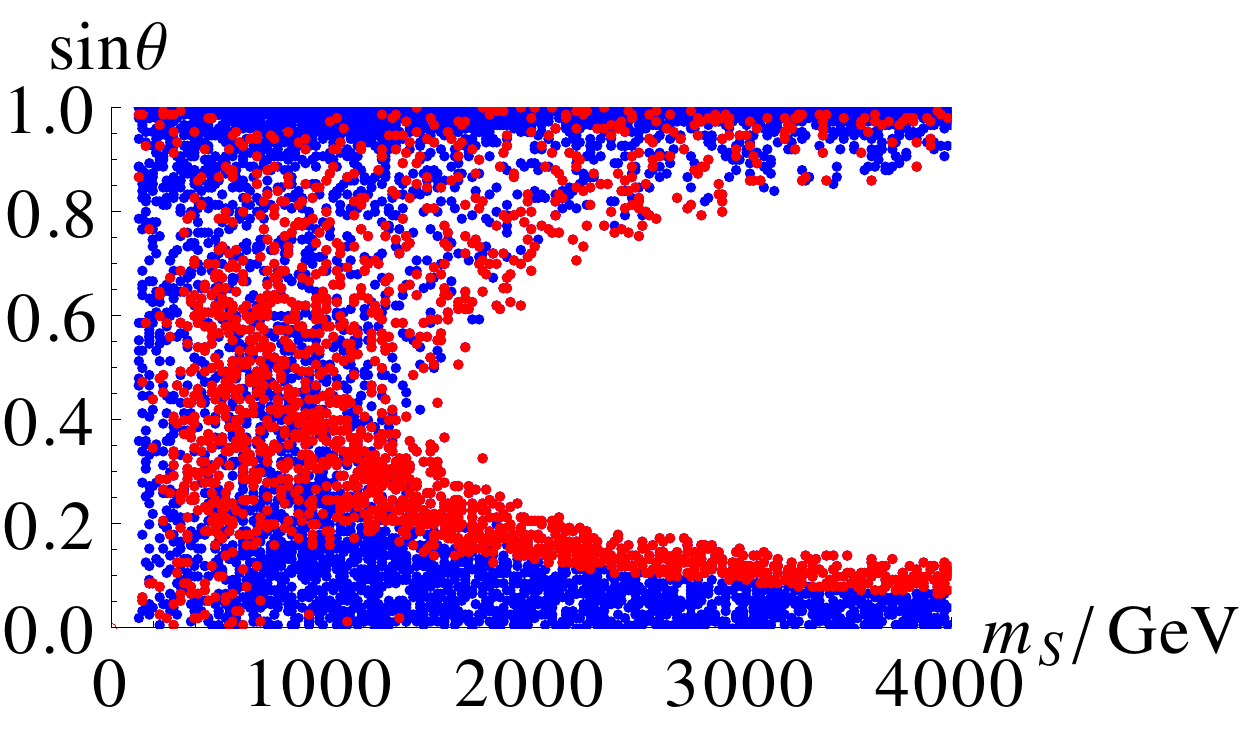}
	\caption{The two classes of phase transitions in the full singlet model. First order phase transitions are shown with red points, while second order transitions correspond to blue points.}
	\label{fig:fullSingletTransitions}
\end{figure}

Several parameters are required for fully specifying a point in the singlet-model parameter space. This makes a description of the region which admits strongly first-order phase transitions somewhat involved. For simplicity, we show the region in Figure~\ref{fig:fullSingletTransitions} by projecting onto the values of the singlet mass and the factor $\sin \theta$ which measures how strongly the singlet couples to the Standard Model. We sample homogeneously in $m_S$, $\delta_2$, $\kappa_3$, $\kappa_4$, and $\theta$, but note that we actually plot versus $\sin\theta$, not $\theta$ itself. 

Some features of Figure~\ref{fig:fullSingletTransitions} are worth discussing. 
First, for large singlet masses (larger than about 1500 GeV) and intermediate values of $\sin \theta$, there is clearly a large region (the white region on the right) completely devoid of physical parameter sets. This was also observed in \cite{Damgaard:2013kva}. In that region, there are (at least) two minima, and the Standard Model minimum is not the global one. For smaller mixing angles ($\sin\theta<0.4$), there is a clear band of red first-order points running along the lower edge of the white region. In this case, there are again two minima, and the Standard Model one is the global minimum at zero temperature. As the temperature increases, the relative heights of the minima change and there is a first-order phase transition. For convenience, we will call this region of first order phase transitions the ``arm''. Below this band is a region of blue second order transition points, where the potential may or may not have more than one minimum, but the global minimum at zero temperature is the Standard Model vacuum. 

The ``arm'' of strongly first-order phase transitions, also observed in~\cite{Damgaard:2013kva}, seems surprising at first. It seems to stretch to arbitrarily large $m_S$, violating the expectation that for large enough $m_S$ the behaviour of the singlet model should be the same as the Standard Model, so that there should only be second order phase transitions. To understand this phenomenon in more detail, we now compare the singlet model more carefully with the effective theory.

\section{Comparing singlet and effective models}
\label{sec:comparison}

Our interest now focuses on the following question: 
to what extent does the effective theory calculation give the correct determination of the order and strength of the transition in the singlet-extended model? That is, under what circumstances does the effective theory provide a reliable estimate of the physics of the phase transition in the singlet model which we take to be the UV completion? To investigate this question, we take a parameter set in the singlet model, and match it to onto the associated effective model using Eqs.~(\ref{eq:match_par}, \ref{eq:match_par_loop}). Then we read off from Fig.~\ref{fig:eff_ranges} whether this parameter set is expected to be first or second order, and compare this to the ``correct" results from the computations in the UV-completed theory, the singlet-extended model.  

\subsection{$Z_2$ symmetry not present}
\label{sec:notz2}

First, we examine the situation where there is no $S \rightarrow -S$ $Z_2$ symmetry. To do so, we must first restrict to points in the singlet model which admit the effective theory: that is, they must have the property that $\kappa_2$ is the only large mass scale in the theory so that our matching relations are valid. We wish to be generous about our choice of region in which we expect the EFT to be valid to gain as broad a coverage as possible, although the cost will be to allow some points in the singlet model parameter space which are somewhat outside the precise scope of the effective theory. Thus, we allow for a violation of the matching relations by a factor of 2:
\begin{align}
\label{eq:crit1}
&\frac12  < 2 \left|\delta Z\right|/\theta ^2 < 2, \\
&\frac{1}{m_h^2}\left[m^2_h -\left( \mu_h^2 - \frac{\mu_{hs}^4}{4 \kappa_2} + \frac{\delta_2 v^2 \mu_{hs}^4 - 
2 \mu_{hs}^4 \mu_h^2}{8 \kappa_2^2} \right)\right] < \frac 12.
\label{eq:crit2}
\end{align}
Further, we require that the quantum mechanical (one loop) corrections to the matching relations are not too large. To achieve this, it is sufficient to impose
\begin{align}
\label{eq:qcInequality}
\frac{\delta_1^2}{8 \kappa_2^2} &> \frac{1}{2} \frac{\delta_2^2}{12 (4 \pi)^2 \kappa_2} .
\end{align}
The final condition we impose is designed to ensure decoupling between the heavy, mostly $S$ state, and the lighter Higgs-like particle. Decoupling requires that interactions between the heavy and light degrees of freedom do not scale with mass of the heavier particle. Again, to be conservative in our choice we allow coupling constants in the Lagrangian which carry dimension of mass to be smaller than $2 \kappa_2^{1/2}$.

\begin{figure}[t]
	\centering
	\includegraphics[width=0.5\textwidth]{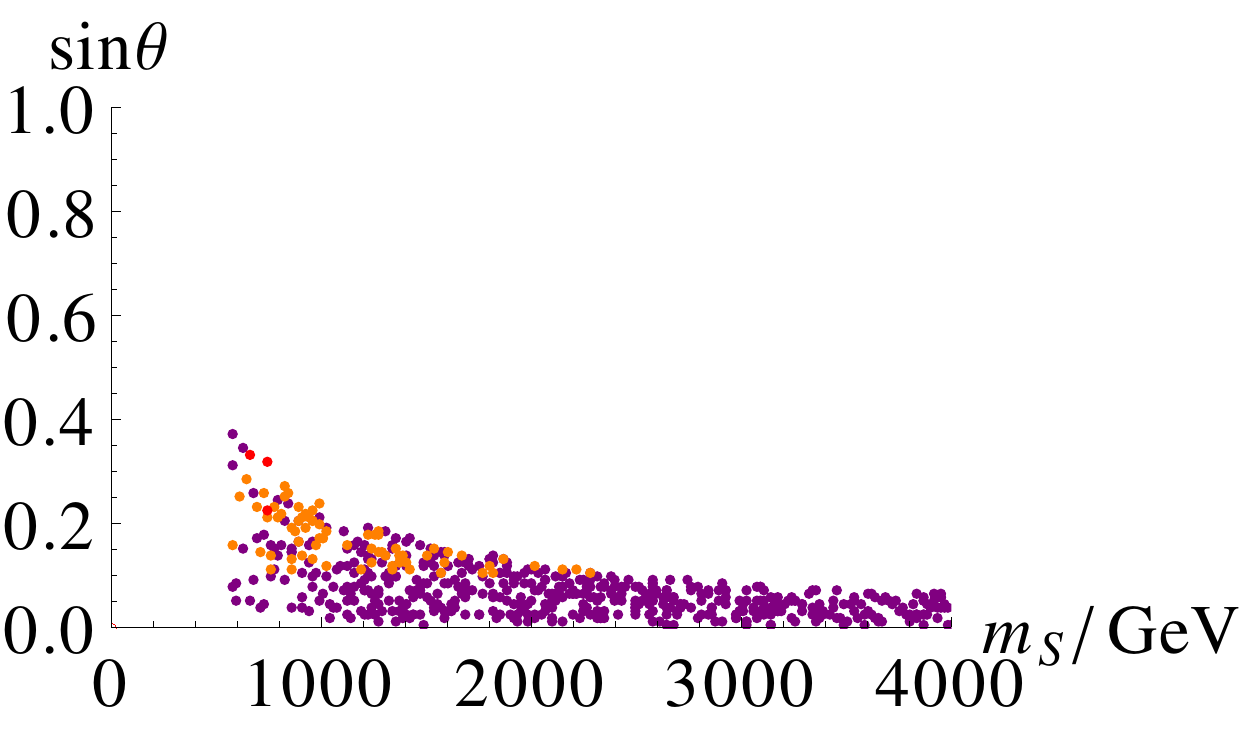}
	\caption{The two classes of phase transitions in the full singlet model, looking only at points where the SM extended by the two dimension 6 operators is a valid description of low-energy physics. First order phase transitions are shown with red points.  Second order transitions correspond to the purple and orange points. The EFT correctly identifies the phase-transition order of the purple points while it incorrectly identifies the order of the phase transition of the orange points.}
	\label{fig:singletEFTscope}
\end{figure}

Points in singlet model parameter space which pass these tests will have perturbative effective field theories which are weakly coupled and can be truncated at dimension 6 to a reasonable approximation. Phase transitions in such singlet models are shown in Figure~\ref{fig:singletEFTscope}.
We project again onto the $m_S$-$\sin\theta$-plane, corresponding to the two quantities most likely to be experimentally accessible. 

Only three points in our scan of singlet model parameter space satisfy our inequalities while generating strongly first-order phase transitions; these are the red points in Figure~\ref{fig:singletEFTscope}. As one would expect, these points have relatively light $m_S \lesssim 1$ TeV. These points also have a strong coupling between the lighter Higgs sector of the theory and the $S$, with mixing angles such that $\sin \theta \gtrsim 0.3$. Indeed, if we require that massive couplings in the scalar theory satisfy the more stringent requirement that they are smaller than $\kappa_2^{1/2}$ (rather than $2\kappa_2^{1/2}$) these points are removed. This confirms the expectation that one requires strong coupling and/or a light new scalar state to change the nature of the Standard Model electroweak phase transition. The ``arm'' we observed in our scan of the full scalar parameter space, Figure \ref{fig:fullSingletTransitions}, corresponds to very large values of the coupling between the Higgs and the $S$, in particular large values of $\delta_1$. Indeed, one needs such large values of $\delta_1$ such that the mixing angle between the Higgs and the $S$ remains comparatively large $\sin \theta \simeq 0.2$ on the ``arm'' for $m_S = 4$ TeV. Although nothing prohibits such large dimensionful couplings that scale with the mass of the heavy state, one can view these cases (and hence the entire "arm" of Figure~\ref{fig:fullSingletTransitions}) as a region of rather fine-tuned parameters. We have probed the region beyond $m_S=4$ TeV and seen the "arm" continue; we take $4$ TeV to approximately represent the likely range of the LHC. We note that $\sin \theta$ is already strongly constrained (see for instance \cite{Damgaard:2013kva}).

The three strongly-first order points which pass our inequalities defining the region where the EFT is approximately valid have the property that both the full singlet theory and its EFT truncation agree on the nature of the phase transition. That is, these points are known to have strongly first-order transitions from an analysis of the full theory. One can construct an EFT for each of these three points, using the matching relations, Eq.~\eqref{eq:match_par}. Then it is straightforward for us to determine the order of the phase transition predicted by the EFT, using our analysis of the EFT phase transition structure as described in Section~\ref{sec:effectivePT}. We find that the EFT predicts all three points have strongly first-order phase transitions, in agreement with the full singlet model.

\begin{figure}[t]
	\centering
	\includegraphics[width=0.5\textwidth]{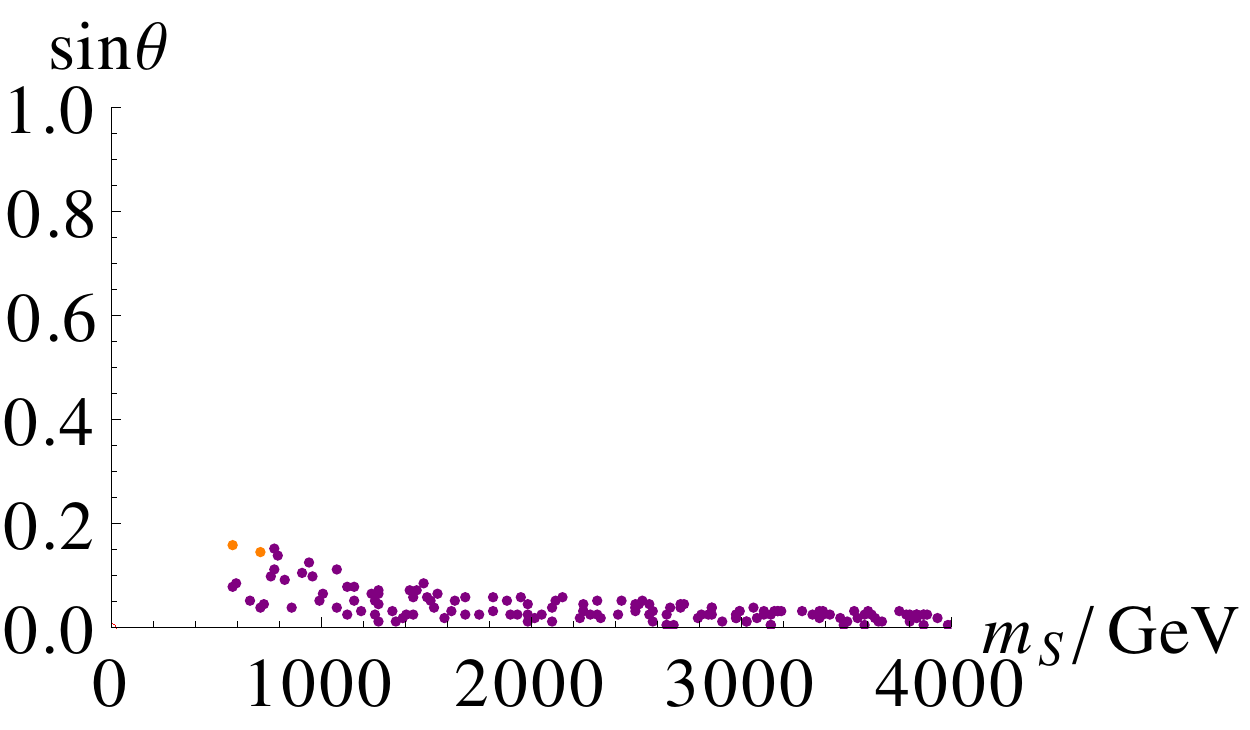}
	\caption{Same as Figure~\ref{fig:singletEFTscope}, except we have reduced the allowed values of dimensionful couplings so that they are less than $\kappa_2^{1/2}$ (rather than $2\kappa_2^{1/2}$.) Notice that only second order phase transitions exists in this region, and furthermore there is more full agreement between the singlet model and the EFT.}
	\label{fig:singletEFTscopeF1}
\end{figure}

Agreement between the effective and full analyses is not guaranteed in our analysis because we have been generous with our definition of the region of validity of the EFT (so that complete agreement is not to be expected). In particular, thermal corrections due to the heavy new physics are taken into account in the full singlet model, while in the EFT, they are not. By being generous about the region of validity of the EFT, we are including regions where the hierarchy of scales defining the EFT is not particularly large, and where coupling are comparatively strong. So deviations are to be expected.
Indeed, we find that the two theories disagree on the nature of the phase transition of a selection of points which the full theory determines have second order phase transitions. Therefore, in Figure~\ref{fig:singletEFTscope}, we color-code points which have second order phase transitions such that purple points have second order transitions according to both the effective and full analyses (agreement.) Meanwhile, the EFT disagrees with the full analysis in the case of the orange points in parameter space. Disagreement between the EFT and the full theory (orange points) occur in the region $m_S \lesssim 2$ TeV and are associated with larger couplings between the heavier and lighter states as evidenced by the substantial mixing angles of orange points. More insight into this region of disagreement can be obtained by strengthening our restriction on values of dimensionful couplings so that they are now smaller than $\kappa_2^{1/2}$ rather than $2\kappa_2^{1/2}$ (Figure~\ref{fig:singletEFTscopeF1}.) As the figure shows, we obtain more detailed agreement between the EFT and the full singlet model in this region, at the expense of completely cutting out strongly first-order phase transitions.
Meanwhile, examination of Figures~\ref{fig:singletEFTscope} and~\ref{fig:singletEFTscopeF1} shows that there is full agreement between the full singlet model and the effective theory for small mixing and large scalar mass, as we would expect. 

In view of the fact that the first-order phase transitions which we find within our EFT region of validity are at the boundary of this region, and are surrounded by points where the full and effective models disagree, we interpret these points as being outside the strict region of validity of the EFT. The agreement between the full and effective theories for these points is presumably coincidental.

\begin{figure}[t]
	\centering
	\includegraphics[width=0.4\textwidth]{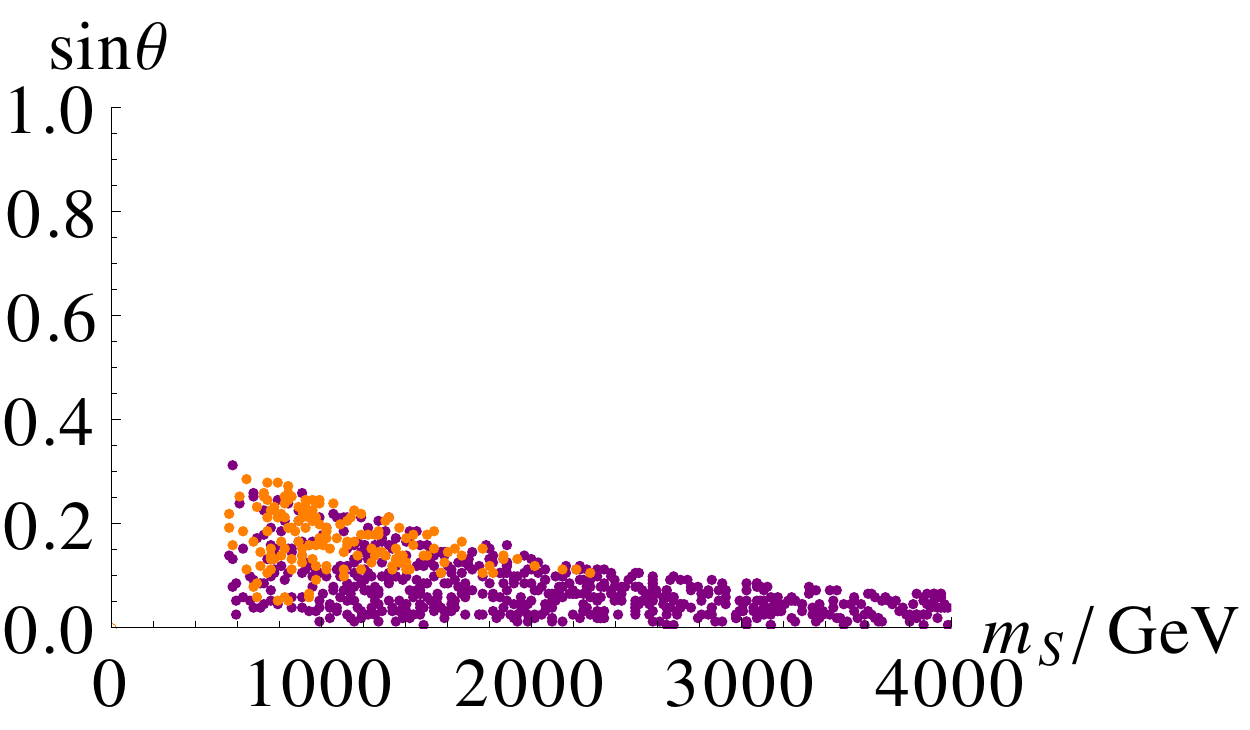}
		\includegraphics[width=0.4\textwidth]{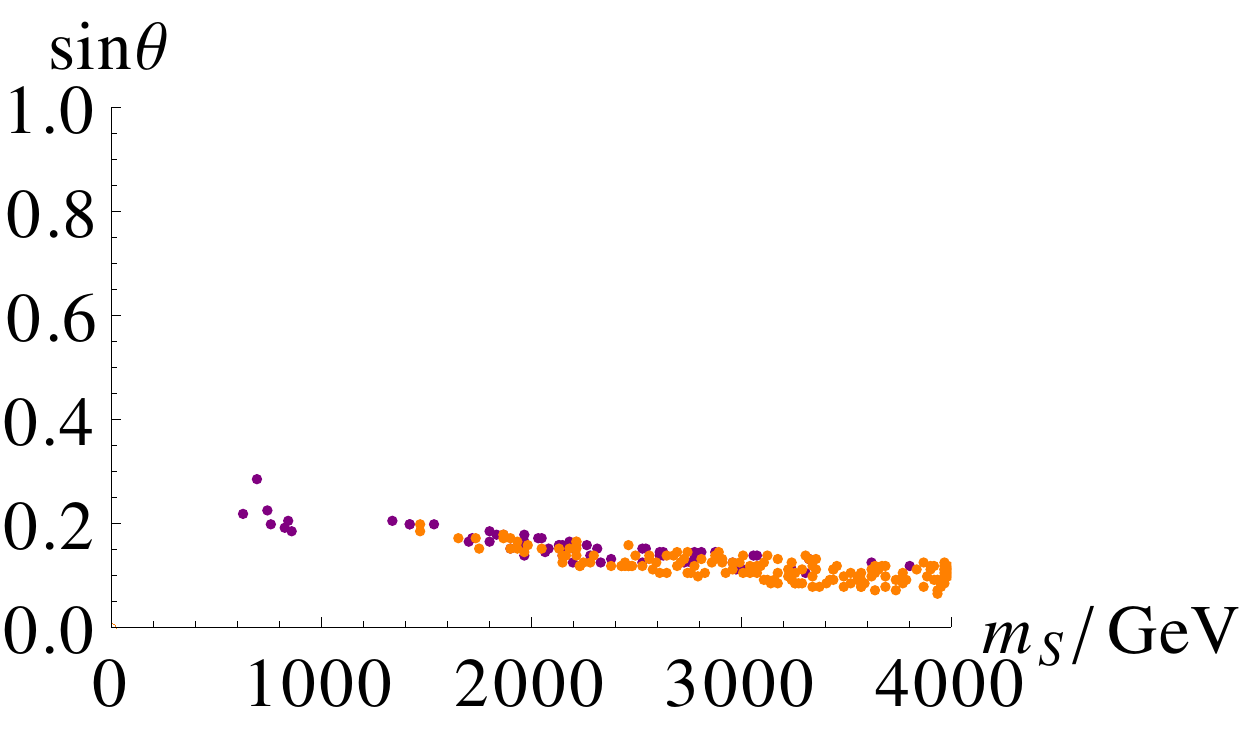}

	\caption{As for Figures \ref{fig:singletEFTscope}, \ref{fig:singletEFTscopeF1} but only restricting dimensionful couplings to be smaller than $4\kappa^{1/2}$ (way beyond the expected validity of EFT). On the left, the points identified as second order in the singlet model; on the right, the points identified as first order in the singlet model. Purple points again refer to where EFT identifies the order correctly. Orange points where it does not.  We see the first order "arm" appearing, which clearly is beyond reach of the EFT treatment. }
	\label{fig:factor4}
\end{figure}

Finally, to further establish the region of validity of the EFT treatment, we repeat the analysis with dimensionful couplings up to $4\kappa_2^{1/2}$, shown in Figure \ref{fig:factor4}. We show the points identified as second order by the singlet model in the left-hand plot, and the points identified as first order in the right-hand plot. Relaxing the cut to $4\kappa_2^{1/2}$ is clearly beyond where we expect EFT to be reliable, and we indeed see that many additional second order points are now orange (and therefore incorrectly identified by EFT). We also see that a large portion of the "arm" of first order points is now included, and is also badly reproduced by EFT. This is as expected, since as we argued the "arm" and the white region beyond correspond to the emergence of a more complicated vacuum structure, as well as large values of dimensionful couplings.

Thus we learn that our effective analysis is only reliable when there is a large separation of scales (including dimensionful couplings.) If one relaxes these requirements, then a richer set of possibilities is present, but a detailed understanding of the physics requires use of the full theory. The effective analysis opens a window only on a relatively small part of the singlet model's parameter space; it is also a region of singlet model parameter space which is less interesting from the point of view of baryogenesis. In particular, the ``arm'' of strongly first-order phase transitions is associated with a large coupling with mass dimension 1. Allowing such a large coupling may seem unnatural, but in view of the fine tuning which is apparently already present in the Standard Model, one should be cautious about a priori discarding such cases.

\subsection{$Z_2$-symmetric theory}
\label{sec:z2}

\begin{figure}[t]
\begin{center}
\includegraphics[width=7cm]{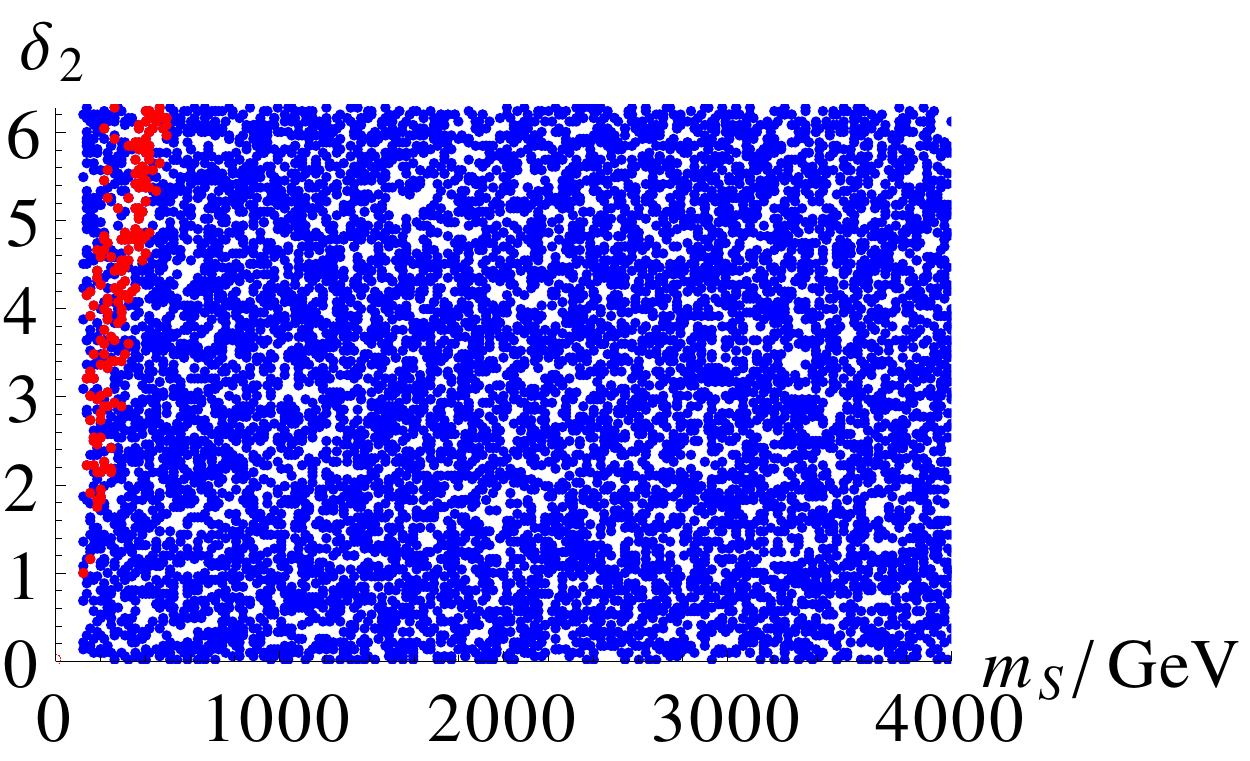}
\caption{The singlet extended model parameter sets, when imposing $Z_2$-symmetry. We find that the red points in parameter space have first order phase transitions, while the blue points have second order transitions.}
\label{fig:sing_overview}
\end{center}
\end{figure}

Having discussed the more generic situation, let us return to the case where we impose a $Z_2$ $S \rightarrow -S$ symmetry on the full singlet theory. In this situation, we choose a slightly different presentation of the parameter space in view of the fact that the mixing angle is exactly zero. In Fig.~\ref{fig:sing_overview}, we project the singlet parameter space $\{\delta_2, m_S, \kappa_4\}$ onto the $\{\delta_2, m_S\}$-plane\footnote{The matching relations are independent of $\kappa_4$ to our order of approximation.}. Points in Fig.~\ref{fig:sing_overview} which are coloured red have strongly first-order phase transitions, while the points coloured blue have second order phase transitions. Notice that the region in parameter space with first-order transitions occurs for fairly small values of $m_S$, below around $500$ GeV with larger values of $m_S$ available for large values of $\delta_2$. This suggests that the region will be outside the domain of validity of our effective analysis, which requires $\kappa_2 \gg m_h^2$. Furthermore, our truncation to the first order of perturbation theory requires not too large values of $\delta_2$. (Notice that we do not impose an analogue of Eq.~\eqref{eq:qcInequality} on the $Z_2$ symmetric theory.)

\begin{figure}[h]
\begin{center}
\includegraphics[width=7cm]{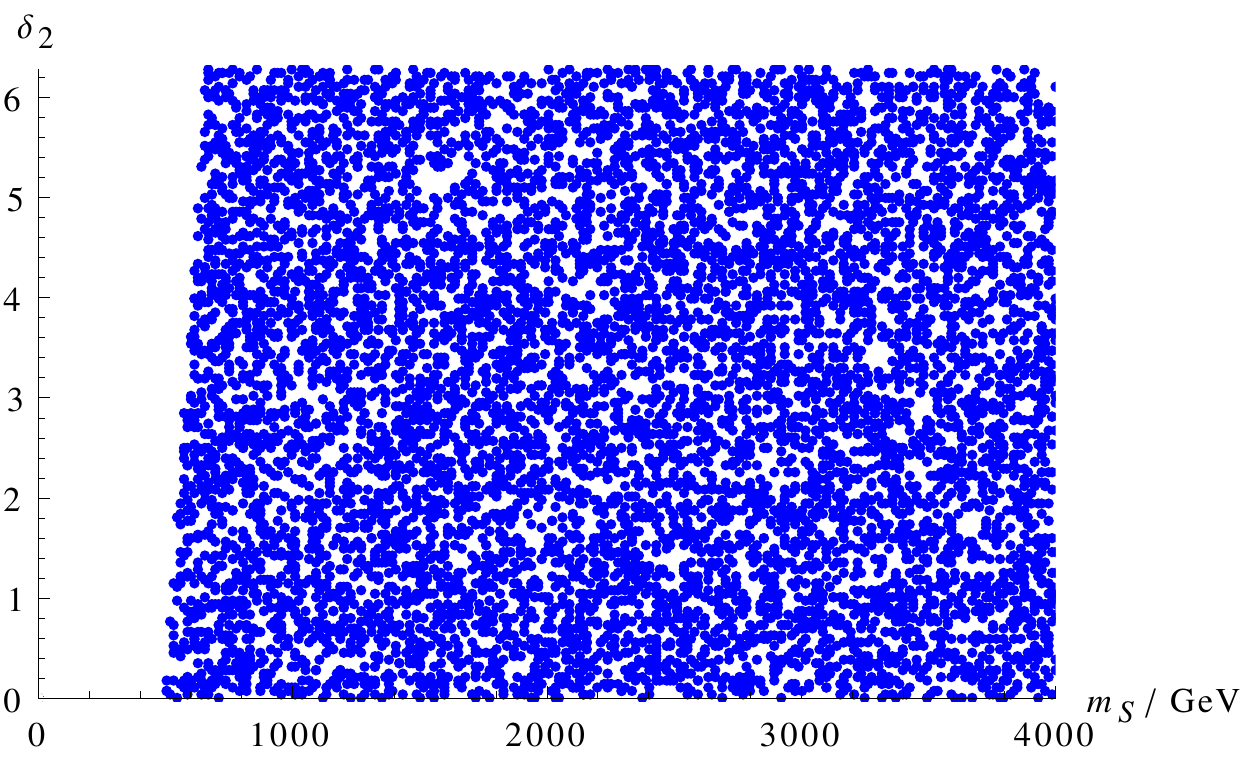}
\includegraphics[width=7cm]{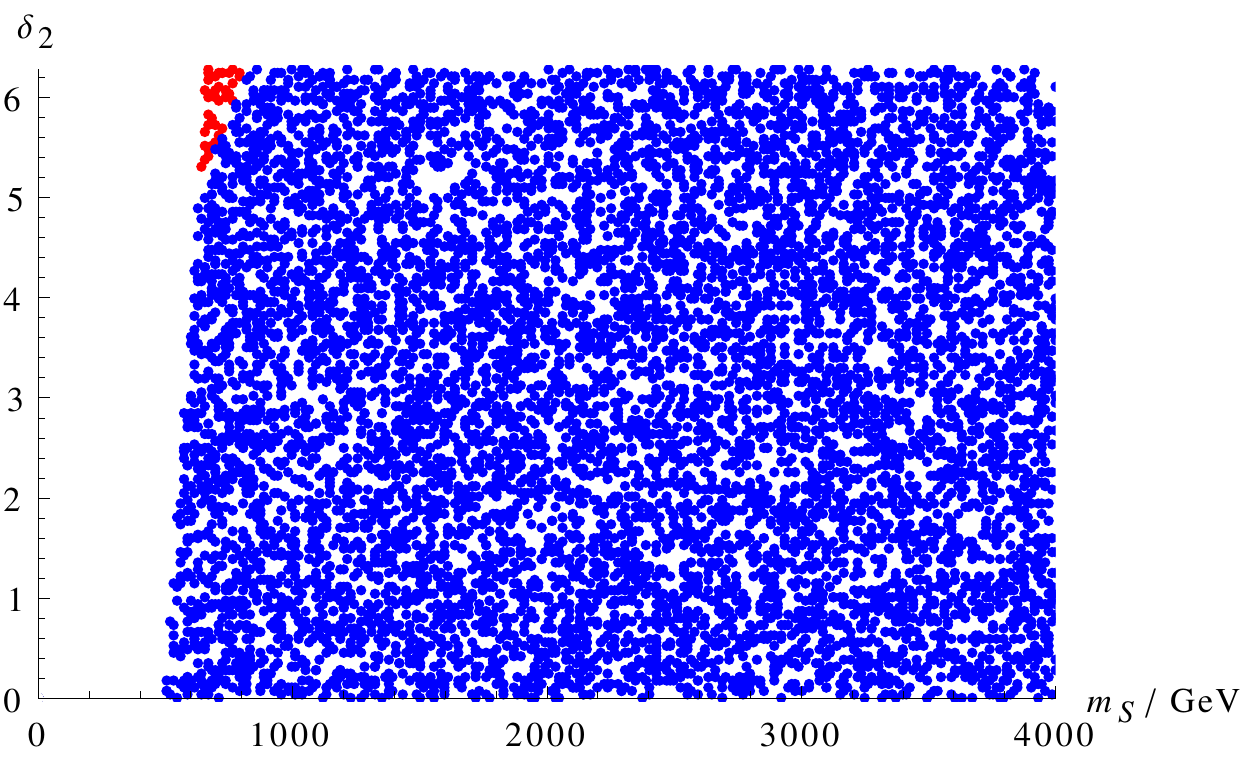}
\caption{The singlet extended model parameter sets, when imposing $Z_2$-symmetry. On the left, transitions in the singlet model, restricting to the approximate region in parameter space where the effective theory is valid. On the right, we colour the same points according to whether the EFT theory predicts first order (red) or second order (blue) phase transitions.}
\label{fig:sing_allinonez2}
\end{center}
\end{figure}

To compare the full and effective theories in more detail, we restrict the parameter space by requiring $\kappa_2^{1/2} > 500$ GeV, so that there is a reasonable hierarchy between the light and heavy mass scales. Again, this is a generous definition of the region of validity of the EFT. We then determine the nature of the phase transition predicted by the effective theory. Our result is shown in Fig.~\ref{fig:sing_allinonez2}. In the left-hand plot, we show the parameter sets that give rise to second order transitions (blue points) and the ones that give rise to first order transitions (red points). In the right-hand plot, we show the same parameter sets, but now we color-code them according to whether their matching EFT parameter set predicts a second order transition (blue points) or a first order transition (red points). 

In the region where we expect the effective theory to be a good description of the physics, the full singlet model predicts only second order phase transitions. In a small corner of the parameter space, at the lower range of $\kappa_2$ and larger $\delta_2$, the effective theory predicts first order phase transitions. In view of the location of this region, we interpret this as a signal that higher order corrections in the EFT have become important. Therefore, once again, we see that the EFT and full singlet models agree, except on the boundary of the region where we expect the EFT to be valid. Moreover, in the core region where the EFT is a good description of the physics, only second-order phase transitions are present.

\section{Conclusions}
\label{sec:conclusions}

It is very tempting to apply EFT to understand the nature of the electroweak phase transition in the context of physics beyond the Standard Model. In this way, one would remain agnostic about the underlying theory but acquire knowledge systematically order-by-order of what this underlying theory could be. To test this approach, we performed a direct comparison between phase transitions in the singlet model and an effective description of the model. This is an important issue since it can teach us the extent to which we can trust EFT predictions regarding electroweak baryogenesis.

One difficulty with the effective approach is the comparatively slow decoupling of thermal effects due to the modest hierarchy between the electroweak scale and the relevant range of singlet scalar masses. This phenomenon occurs for example in our analysis of second order phase transitions, shown in Fig.~\ref{fig:singletEFTscope}. For $500$ GeV $\lesssim m_S \lesssim 2000$ GeV and mixing angles $\theta \gtrsim 0.1$, our EFT description of the phase transition structure is incorrect, in that there is a mismatch with the phase transition structure of the full singlet model. This is surprising at first, since the separation between physical states is almost an order of magnitude. However, couplings in this region are fairly large, and the exponential Boltzmann suppression does not yet dominate.

For larger masses $m_S > 2$ TeV or smaller mixing angles, there is a match between the effective and singlet models. This is the decoupling region in accordance with expectation. However, the full singlet model has a much richer structure of phase transitions which are associated with stronger couplings between the light (mostly Higgs) sector and the heavy new physics sector as seen in Figure~\ref{fig:fullSingletTransitions}. While the coupling between Higgs and $S$ is quite large, for example on the ``arm'' of strongly first-order phase transitions, it does not appear to be ruled out by any issue of principle. The singlet theory itself is still perturbatively coupled in this region. But there is no decoupling between the light and heavy modes due to a large dimensionful coupling. Therefore, the effective analysis is limited to a fairly small region of singlet model parameter space, which also happens to be a region which is less interesting from the point of view of baryogenesis. We have discussed the extent to which this particular region of parameter space can be dismissed as fine-tuned.

We further investigated imposing a $Z_2$ symmetry on the model. In this case, we find that within the domain of validity of the EFT there is broad agreement between the full and effective analysis of the phase transitions. In this region, phase transitions are second order. One needs to couple the new physics more strongly to the Higgs sector in order to generate strongly first-order phase transitions.

Thus we conclude that an effective analysis of baryogenesis must be treated with some caution. A modest hierarchy of scales can lead to situations in which an EFT analysis disagrees with a more detailed analysis in the full UV completion. Moreover, detailed UV theories can easily contain strongly first-order phase transitions which are far away in parameter space from the region covered by effective theory. 

\section*{Acknowledgements}\label{sec:Acknowledgements} 
We thank Richard Ball, Simon Badger, Arjun Berera and Michael Trott for useful discussions. AT acknowledges support from the Villum Foundation. PHD and DOC thank the KITP at Santa Barbara for hospitality while this work was being completed. This research has been partly supported by the National Science Foundation under Grant No. NSF PHY11-25915, the STFC consolidated grant ``Particle Physics at the Higgs Centre'', and by the Marie Curie FP7 grant 631370.

\appendix

\section{Effective potential contributions, effective theory}
\label{app:finiteT_EFT}

\subsection{EFT: Tree-level potential $V$}
\label{app:Clas_effective}

The classical potential of the EFT model is
\beq
V(H)=- \frac{\lambda_4}{4} \left(H^\dagger H - \frac{v^2}{2} \right)^2 - \frac{\lambda_6}{\Lambda^2} \left(H^\dagger H - \frac{v^2}{2} \right)^3.
\eeq
\\
Keeping in mind the presence of the kinetic dimension 6 operator, we write again $H =\frac{1}{\sqrt{2}} (\phi + h)$, where $\phi$ is a constant background field and $h$ the fluctuations around $\phi$, we define $h^n =\tilde h^n(1-n\delta Z)$. Then $\tilde h$ is canonically normalized. 

We have
\ba
V(\phi, \tilde{h}) &=& V(\phi) + V_{\tilde{h}}(\phi,\tilde{h})=\nonumber\\&&
- \left(\frac{\lambda_4 v^2}{8} -\frac{3v^4 \lambda_6}{8 \Lambda^2}\right)\phi ^2 
+\left(\frac{\lambda_4}{16}  - \frac{3 \text{$\lambda_6$} v^2}{8 \Lambda ^2}\right) \phi ^4 
+   \frac{\text{$\lambda_6$}}{8 \Lambda ^2}\phi^6\nonumber\\&&
+(1- \delta Z) \bigg(\frac{\lambda_4}{4}(\phi^3 - v^2 \phi)
+ \frac{3\lambda_6}{4 \Lambda^2} (\phi^5 + v^4 \phi - 2 v^2 \phi^3) \bigg)\text{$\tilde h$}\nonumber\\&&
+ (1-2 \delta Z) \bigg( \frac{\lambda_4}{8}(3\phi^2 - v^2)
+ \frac{\lambda_6}{8 \Lambda^2} (15 \phi^4 +3 v^2 - 18v^2 \phi^2)\bigg)\text{$\tilde h$}^2\nonumber\\&&
+ (1-3 \delta Z)\left(\frac{\lambda_4}{4}\phi + \frac{\lambda_6}{2\Lambda^2} (5 \phi^2 - 3 v^2)\phi \right)\text{$\tilde h$}^3\nonumber\\&&
+  (1-4\delta Z)\left( \frac{\lambda_4}{16}  + \frac{3\lambda_6}{8\Lambda^2}(5 \phi^2-v^2) \right)\tilde h^4\nonumber\\&&
+ (1-5\delta Z) \frac{3 \text{$\lambda_6$} \phi }{4 \Lambda ^2} \tilde h^5
+ (1-6\delta Z)\frac{\text{$\lambda_6$}}{8 \Lambda ^2} \tilde h^6+ \frac{\lambda_4}{16}v^4 - \frac{\text{$\lambda_6$} v^6}{8 \Lambda ^2},
\ea
where the first line is the classical potential $V$ and the last two terms are constants that have no consequence for the location for the potential minimum. By differentiation with respect to $\phi$, we find that $v$ is the minimum of the potential, as required. By differentiation with respect to $\tilde{h}$ twice, we identify the pole mass at $\phi=v$ for the degree of freedom $\tilde{h}$ to be $\tilde{m}^2=(1-2 \delta Z)\frac{\lambda_4v^2}{2}$ (see also Eq.~\ref{higgsmass}). From higher derivatives with respect to $\tilde{h}$ we find the 3-, 4-, 5-, and 6- point vertices to be
\beq
&\frac{v_3}{3!} = (1 - 3 \delta Z) \left(\frac{\lambda v}{4} + \lambda_6 \frac{v^3}{\Lambda^2}\right), \\
&\frac{v_4}{4!} = (1 - 4 \delta Z)\left( \frac{\lambda}{16} + \frac{3\lambda_6}{2} \frac{v^2}{\Lambda^2}\right), \\
&\frac{v_5}{5!}= (1 - 5 \delta Z) \frac{3\lambda_6v}{4\Lambda^2}, \\
&\frac{v_6}{6!} = (1 - 6 \delta Z)  \frac{\lambda_6\Lambda^2}{8} .
\eeq

\subsection{EFT: The Coleman-Weinberg potential $V_{CW}$}
\label{app:CW_effective}

After divergences have been subtracted, the zero-temperature contribution at one-loop order to the effective potential has the form 
\beq
V_{\rm CW}=\sum_{i}\frac{1}{64\pi^2} N_i M_i^4(\phi)\left[\log\frac{M_i^2(\phi)}{\mu^2}-C_i\right],
\eeq
where $\mu$ is an arbitrary scale, $i$ sums over particle species in the Standard Model and
\beq
\label{eq:theNs}
N_{t,c,u,d,s,b}=-12,\quad N_{W}=6,\quad N_Z=3,\\
\quad N_{h}=1,\quad N_{G}=3,\quad N_{e,\mu,\tau,\nu_e,\nu_\mu,\nu_\tau}=-4.\nonumber
\eeq
We have ignored the photon and the gluons, since they do not couple directly to the Higgs field. Their contribution would therefore only be a constant as a function of $\phi$, and hence is irrelevant to the location of the minimum. We have split the four Higgs degrees of freedom into the massive mode $h$ and the three massless modes $G$. $C_i$ is 5/6 for gauge bosons, 3/2 for the rest. $\mu$ is a renormalization scale which we take to be $m_t$.

The zero temperature masses as a function of the Higgs and singlet fields are:
\beq
M_{t,c,u,d,s,b,W,Z,e,\mu,\tau,\nu_e,\nu_\mu,\nu_\tau}(\phi)=m_{t,c,u,d,s,b,W,Z,e,\mu,\tau,\nu_e,\nu_\mu,\nu_\tau}\frac{\phi}{v}.
\eeq
The $\phi$-dependent  mass eigenvalue of the Higgs field fluctuations is the second derivative of  $V(\phi,\tilde{h})$ with respect to $\tilde h$:
\beq
M_h^2(\phi) = (1-2 \delta Z) \bigg( \frac{\lambda_4}{4}(3\phi^2 - v^2)
+ \frac{3\lambda_6}{4 \Lambda^2} (5 \phi^4 + v^2 - 6v^2 \phi^2)\bigg).
\eeq
The $\phi$-dependent mass of the massless modes follows from inserting the doublet of fluctuations 
\ba
H\rightarrow \left(\begin{array}{c}
\chi_1+i \chi_2\\
\frac{1}{\sqrt{2}}(\phi+h)+i\chi_3,
\end{array}\right)
\ea
canonically normalized into the Lagrangian. The $\phi$-dependent mass is now given by the second derivative with respect to $\tilde{\chi}_{1,2,3}$. These are all the same and read\beq
M_G^2(\phi) = (1-2 \delta Z) \bigg( \frac{\lambda_4}{4}(\phi^2 - v^2)
+ \frac{3\lambda_6}{4 \Lambda^2} (\phi^2-v^2)^2\bigg).
\eeq
The mass of transverse (Goldstone) modes vanishes in the zero temperature vacuum $\phi=v$, as it must.

\subsection{EFT: Counterterms $V_{ct}$}
\label{app:CT_effective}

In order to make the one-loop contribution finite, we introduce counterterms,
\beq
V_{\rm ct}= \delta V_0+ \frac{1}{2}\delta m^2 H^\dagger H + \frac{1}{4}\delta\lambda (H^\dagger H)^2.\nonumber
\eeq
These cancel all divergences and we fix their finite parts by imposing three renormalization conditions. 
\beq
(V_{ct} + V_{CW})\bigg{|}_{v} = 0,\qquad
\frac{\partial (V_{ct} + V_{CW})}{\partial H^\dagger} \bigg{|}_{v} = 0,\qquad 
\frac{\partial^2 (V_{ct} + V_{CW})}{\partial H\partial H^\dagger} \bigg{|}_{v} = 0.
\eeq
The renormalization conditions amount to enforcing that the position and depth of the minimum and the mass at one-loop order are the same as they are at tree-level. 
In the renormalisation procedure, we have ignored the ``Goldstone" contributions, as their derivatives are badly behaved. Since the contribution to the potential vanishes in the vacuum (their mass is zero) and is small near the minimum, computing the counter terms without, and then computing the potential with, amounts to a very small error \cite{Cline:2011mm}. 

\subsection{EFT: Finite-temperature contribution $V_T$}
\label{app:finiteT_effective}
At one-loop order we write the finite-temperature contribution as
\beq
V_T(\phi,T) = V_T^1(\phi,T) + V_T^{\text{ring}}(\phi,T).
\eeq
The first component is the one-loop expression
\beq
\label{oneloopT}
V_T^1(\phi,T) = \frac{T^4}{2\pi^2} N_i \int_0^{\infty} dx x^2 \text{log}\bigg[ 1 \pm e^{-\sqrt{x^2 + \frac{M_i^2(\phi)}{T^2}}} \bigg],
\eeq
where $N_i$ is as above and $M_i(\phi)$ is the zero-temperature field-dependent mass. The $\pm$ refers to fermions and bosons respectively. Expanding this for small $\frac{M_i}{T}$ yields \cite{Arnold:1992rz}:
\begin{align}
&\text{const.} + \frac{1}{24} N_i M_i^2(\phi) T^2 - N_i \frac{T}{12 \pi} M_i^3(\phi) + O(M_i^4), &\text{(bosons)}
\label{expansion-bosons} \\
&\text{const.} + \frac{1}{48} N_i M_i^2(\phi)T^2 + O(M_i^4), &\text{(fermions)}.
\label{expansion-fermions}
\end{align}
There are now two procedures for including thermal corrections to the effective masses. One is to follow \cite{Cline:1996mga} by simply replacing
\ba
M_i^2(\phi)\rightarrow M_i^2(\phi,T),
\ea
in (\ref{oneloopT}). The other follows \cite{Arnold:1992rz},\cite{Carrington}  and involves making the exchange
\ba
M_i^3(\phi)\rightarrow M_i^3(\phi,T),
\ea
in (\ref{expansion-bosons}). This amounts to the daisy resummation, and only involves bosonic degrees of freedom. To leading order, we have
\beq
\label{eq:ring1}
V_T^{\text{ring}}(\phi,T) = \sum_i \frac{T}{12\pi} N_i \text{Tr}\bigg[M_i^3(\phi) - M_i^3(\phi,T) \bigg].
\eeq
Note that this effectively swaps the cubic mass term from the above expansion with a thermally dependent one. For either implementation, we need the thermally corrected masses $M_i(\phi,T)$ for each bosonic degree of freedom. Here both the Higgs and Goldstone (The 3 Higgs-field modes orthogonal to $\phi$) modes contribute, and we have a contribution from the gauge-bosons. We have for Goldstone modes
\begin{align}
M_G^2(\phi,T) =& M_G^2(\phi) ~~+ \\   &\bigg[ \frac{3}{16} g^2 + \frac{1}{16} g'^{2} + \bigg(\frac{1}{4} \sum_{\text{quarks}}Y_i^2 + \frac{1}{12} \sum_{\text{leptons}}Y_j^2 + \frac{1}{2}\lambda \bigg) \bigg]T^2 \nonumber
\end{align}
Note that at $v=0$ the first term vanishes and at $T=0$ the second term vanishes. This reflects that the Goldstone modes acquire mass from two separate mechanisms. One corresponds to the Higgs field taking expectation values other than the electroweak minimum and the other from thermal corrections.
For the massive Higgs mode
\begin{align}
M_h^2(\phi,T)=& M_h^2(\phi)~~+ \\   &\bigg[ \frac{3}{16} g^2 + \frac{1}{16} g'^{2} + \bigg(\frac{1}{4} \sum_{\text{quarks}}Y_i^2 + \frac{1}{12} \sum_{\text{leptons}}Y_j^2 + \frac{1}{2}\lambda \bigg) \bigg]T^2 \nonumber
\end{align}
where we have defined the Yukawa coupling constants as:
\begin{align}
\sum_{\rm quarks}Y_i^2&=Y_t^2+Y_b^2+Y_c^2+Y_s^2+Y_d^2 + Y_u^2 , \\ \qquad \sum_{\rm leptons}Y_j^2&=Y_e^2+Y_\mu^2+Y_\tau^2+Y_{\nu_e}^2+Y_{\nu_\mu}^2+Y_{\nu_\tau}^2.\nonumber 
\end{align}
For the gauge bosons we revert to the original gauge field basis, and write the mass matrix
\beq
M^2(\phi,T)& =& M^2(\phi) + M_T^2(T)=\nonumber\\
&&\left(\begin{array}{cccc}
g^2\phi^2/4&0&0&0\\
0&g^2\phi^2/4&0&0\\
0&0&g^2\phi^2/4&-gg'\phi^2/4\\
0&0&-gg'\phi^2/4&g'^{2}\phi^2/4\end{array}\right)
+\left(\begin{array}{cccc}
\frac{11}{6}g^2T^2&0&0&0\\
0&\frac{11}{6}g^2T^2&0&0\\
0&0&\frac{11}{6}g^2T^2&0\\
0&0&0&\frac{11}{6}g'^2T^2\\
\end{array}\right),\nonumber\\
\eeq
and here the trace in (\ref{eq:ring1}) becomes relevant
\beq
\text{Tr}[M^3(\phi)-M(\phi,T)^3] &= \text{Tr}[M^3(\phi)] - \text{Tr}[M^3(\phi,T)] \\
&= \text{Tr}[D^3_{M(\phi)}] - \text{Tr}[D^3_{M(\phi,T)}].
\eeq
Note that diagonalizing $D^3_{M(\phi)}$ is the same as in the Standard Model, but when diagonalizing $D^3_{M(\phi,T)}$, the $Z$ and $\gamma$ mixes because of the photons thermal mass. This correction makes the longitudinal parts of the gauge-boson fields temperature dependent. It does not correct the transverse parts.   
\\

\section{Effective potential contributions, singlet extended model}
\label{app:singleteff}
 
\subsection{Singlet model: Coleman-Weinberg potential $V_{\rm CW}$}
\label{app:CW_singlet}
When including the singlet field the one-loop zero-temperature contributions are similar
\beq
V_{\rm CW}=\sum_{i}\frac{1}{64\pi^2} N_i M_i^4(\phi,s)\left[\log\frac{M_i^2(\phi,s)}{\mu^2}-C_i\right],
\eeq
where, again, $\mu$ is an arbitrary scale while $i$ now sums over particle species in the Standard Model plus the singlet, and
\beq
\label{eq:theNstoo}
&N_{t,c,u,d,s,b}=-12,\quad N_{W}=6,\quad N_Z=3, \\ \nonumber\quad &N_{s,h}=1,\quad N_{G}=3,\quad N_{e,\mu,\tau,\nu_e,\nu_\mu,\nu_\tau}=-4,
\label{multiplicitySM+singlet}
\eeq
and we have once more ignored the photon and the gluons. We have split the four Higgs degrees of freedom into the massive mode $h$ and the three Goldstone modes $G$. $C_i$ is 5/6 for gauge bosons, 3/2 for the rest. $Q$ is a renormalization scale which we take to be $m_t$.
\\
\\
The zero temperature masses as a function of the Higgs and singlet fields are
\beq
M_{t,c,u,d,s,b,W,Z,e,\mu,\tau,\nu_e,\nu_\mu,\nu_\tau}(\phi,s)=m_{t,c,u,d,s,b,W,Z,e,\mu,\tau,\nu_e,\nu_\mu,\nu_\tau}\frac{\phi}{v},
\eeq
and the field dependent eigenvalues of the Higgs-singlet mass matrix 
\beq
M^2(\phi,s)=\left (
\begin{array}{cc}
\frac{m^2}{2}+\frac{3\lambda}{4}\phi^2+\frac{\delta_1}{2}s+\frac{\delta_2}{2}s^2
&  
\frac{\delta_1}{2}\phi+\delta_2 s\phi
\\
\frac{\delta_1}{2}\phi+\delta_2 s\phi
& 
\kappa_2+\frac{\delta_2}{2}\phi^2+ 2\kappa_3s+3\kappa_4s^2
\end{array} \nonumber
\right ),
\eeq
and the Goldstone modes
\beq
M_G^2(\phi)=\frac{m^2}{2}+\frac{\lambda}{4}\phi^2+\frac{\delta_1}{2}s+\frac{\delta_2}{2}s^2.
\eeq

\subsection{Singlet model: Counterterms $V_{\rm ct}$}
\label{app:CT_singlet}

In order to make the one-loop contribution finite, we introduce a set of counterterms. 
\beq
V_{\rm ct}=\delta V_0+\frac{1}{2}\delta m^2H^\dagger H+\frac{1}{4}\delta\lambda (H^\dagger H)^2+\frac{1}{4}\delta\delta_1S H^\dagger H+\frac{1}{2}\delta\kappa_2 S^2\nonumber\\
+\frac{1}{4}\delta\delta_2S^2 H^\dagger H+\delta\kappa_1 S+\frac{1}{3}\delta\kappa_3S^3+\frac{1}{4}\delta\kappa_4S^4,
\eeq
and we may again insert $H=\frac{1}{\sqrt{2}}(\phi+h)$.
These counterterms cancel all divergences and we fix their finite parts by imposing a set of 9 renormalization conditions. We again ignore the contributions to $V_{\rm CW}$ from the Goldstone modes. 
\beq
(V_{\rm CW}+V_{\rm ct})_{v,0}=
\frac{\partial(V_{\rm CW}+V_{\rm ct})}{\partial \phi}_{v,0}=
\frac{\partial(V_{\rm CW}+V_{\rm ct})}{\partial s}_{v,0}=0,\\
\frac{\partial^2(V_{\rm CW}+V_{\rm ct})}{\partial h^2}_{v,0}=
\frac{\partial^2(V_{\rm CW}+V_{\rm ct})}{\partial s^2}_{v,0}=
\frac{\partial^2(V_{\rm CW}+V_{\rm ct})}{\partial h\partial s}_{v,0}=0,\\
\frac{\partial^4(V_{\rm CW}+V_{\rm ct})}{\partial s^2\partial\phi^2}_{v,0}=
\frac{\partial^3(V_{\rm CW}+V_{\rm ct})}{\partial s^3}_{v,0}=
\frac{\partial^4(V_{\rm CW}+V_{\rm ct})}{\partial s^4}_{v,0}=0.
\eeq
We note that this is explicitly different from the approach in \cite{Espinosa:2011ax,Damgaard:2013kva}, where the renormalisation conditions are a mixture of constraints in the broken and symmetric phase. This leads to either divergent counterterms in the $Z_2$-symmetric limit, or an explicit breaking of $Z_2$ symmetry through the renormalisation conditions, even when $\sin\theta=\kappa_3=\kappa_1=\delta_1=0$. hence our renormalised theory is quite different form the one presented there.

\subsection{Singlet model: Finite temperature contribution $V_{T}$}
\label{app:finiteT_singlet}

The finite temperature contribution are
\beq
V_T(\phi,s,T)=V_T^1(\phi,s,T)+V_T^{\rm ring}(\phi,s,T).
\eeq

The first component is the one-loop expression
\beq
V_T^1(\phi,s,T)=\frac{T^4}{2\pi^2}N_i\int_0^\infty dx\,x^2\,\log\left[1\pm e^{-\sqrt{x^2+M_i^2(\phi,s)/T^2}}\right].
\eeq
where $M_i(\phi,s)$ is the zero-temperature masses already defined. The sign $\pm$ refers to fermions and bosons respectively.
\\
\\
Again we have the daisy resummation:
\beq
V_T^{\rm ring}(\phi,s,T)=\sum_i\frac{T}{12\pi}N_i\textrm{Tr}\left[M^3_i(\phi,s)-M_i^3(\phi,s,T)\right],
\label{eq:ring}
\eeq
where $M^2_i(\phi,s,T)$ refers to the thermally corrected masses, for the 3 Goldstone modes, 
\beq
M_G^2(\phi,s,T)&=&M_G^2(\phi) +\\ &&\bigg[ \frac{3}{16} g^2 + \frac{1}{16} g'^{2} + \bigg(\frac{1}{4} \sum_{\text{quarks}}Y_i^2 + \frac{1}{12} \sum_{\text{leptons}}Y_j^2 + \frac{1}{2}\lambda \bigg) \bigg]T^2,\nonumber\\ \nonumber
\eeq
and for the Higgs and singlet modes
\beq
M^2(\phi,s,T)&=&M^2(\phi,s) +\\ &&\left(\begin{array}{cc}
\bigg[ \frac{3}{16} g^2 + \frac{1}{16} g'^{2} + \bigg(\frac{1}{4} \sum_{\text{quarks}}Y_i^2 + \frac{1}{12} \sum_{\text{leptons}}Y_j^2 + \frac{1}{2}\lambda \bigg) \bigg]T^2&0\\
0&(8\lambda_m+12\lambda_s)T^2
\end{array}\right),\nonumber\\ \nonumber
\eeq
where we have defined the Yukawa coupling constants as:
\begin{align}
\sum_{\rm quarks}Y_i^2&=Y_t^2+Y_b^2+Y_c^2+Y_s^2+Y_d^2 + Y_u^2 , \\ \qquad \sum_{\rm leptons}Y_j^2&=Y_e^2+Y_\mu^2+Y_\tau^2+Y_{\nu_e}^2+Y_{\nu_\mu}^2+Y_{\nu_\tau}^2.\nonumber 
\end{align}
There is also a contribution to $V_T^{\rm ring}$ from the gauge bosons, and for this we have to revert to the original gauge field basis, and write the mass matrix
\beq
\mathcal{M}^2(\phi,s,T)=\left(\begin{array}{cccc}
g^2\phi^2/4&0&0&0\\
0&g^2\phi^2/4&0&0\\
0&0&g^2\phi^2/4&-gg'\phi^2/4\\
0&0&gg'\phi^2/4&g'^{2}\phi^2/4
\end{array}\right)
+\left(\begin{array}{cccc}
\frac{11}{6}g^2T^2&0&0&0\\
0&\frac{11}{6}g^2T^2&0&0\\
0&0&\frac{11}{6}g^2T^2&0\\
0&0&0&\frac{11}{6}g'^{2}T^2\\
\end{array}\right),\nonumber\\
\eeq
and here the trace in (\ref{eq:ring}) becomes relevant.

 
\end{document}